\documentclass[a4paper,fleqn,usenatbib]{mnras}

\usepackage[T1]{fontenc}
\usepackage{ae,aecompl}

\usepackage{graphicx}	
\usepackage{amsmath}	
\usepackage{amssymb}	
\usepackage{natbib}
\setcitestyle{comma, numbers, sort&compress, super}
\usepackage{siunitx}
\usepackage{multicol}

\newcommand*{\rev}{\color{black}}

\usepackage{enumitem}
\usepackage{newtxtext,newtxmath}

\title[Characterising Polarization in Nearby Molecular Clouds]{Characterising the Magnetic Fields of Nearby Molecular Clouds using Submillimeter Polarization Observations}

\author[C. Sullivan et al.]{
Colin H Sullivan,$^{1,2}$\thanks{E-mail: colin.sullivan@virginia.edu}
L. M. Fissel,$^{2,3}$
P. K. King,$^{1,4,5}$
C.-Y. Chen,$^{1,4}$
Z.-Y. Li,$^{1}$
and J. D. Soler$^{6}$
\\
\\
$^{1}$Department of Astronomy, University of Virginia, Charlottesville VA, 22904\\
$^{2}$National  Radio  Astronomy  Observatory,  Charlottesville,  VA,  22904\\
$^{3}$Department of Physics, Engineering Physics, and Astronomy, Queen's University,  Kingston,  ON, Canada, K7L 3N6\\
$^{4}$Johns Hopkins University Applied Physics Laboratory, Laurel, MD, 20723\\
$^{5}$Lawrence Livermore National Laboratory, Livermore, 7000 East Ave, Livermore, CA 94550, USA\\
$^{6}$Max Planck Institute for Astronomy, K{\"o}nigstuhl 17, 69117 Heidelberg, Germany\\
}

\date{Accepted 2021 February 23. Received 2021 February 21; in original form 2019 October 30}

\begin{document}
\label{firstpage}
\pagerange{\pageref{firstpage}--\pageref{lastpage}}
\maketitle

\begin{abstract}
Of all the factors that influence star formation, magnetic fields are perhaps the least well understood. The goal of this paper is to characterize the 3D magnetic field properties of nearby molecular clouds through various methods of statistically analyzing maps of polarized dust emission. Our study focuses on nine clouds, with data taken from the \textit{Planck} Sky Survey as well as data from the BLASTPol observations of Vela\,C. We compare the distributions of polarization fraction (\textit{p}), dispersion in polarization angles ($\mathcal{S}$), and hydrogen column density ($N_\textrm{H}$) for each of our targeted clouds. To broaden the scope of our analysis, we compare the distributions of our clouds' polarization observables with measurements from synthetic polarization maps generated from numerical simulations. We also use the distribution of polarization fraction measurements to estimate the inclination angle of each cloud's cloud-scale magnetic field. We obtain a range of inclination angles associated with our clouds, varying from $16^\circ$ to $69^\circ$. We establish inverse correlations between $p$ and both $\mathcal{S}$ and $N_{\rm H}$ in almost every cloud, but we are unable to establish a statistically robust $\mathcal{S}$ vs $N_{\rm H}$ trend. By comparing the results of these different statistical analysis techniques, we are able to propose a more comprehensive view of each cloud's 3D magnetic field properties. These detailed cloud analyses will be useful in the continued studies of cloud-scale magnetic fields and the ways in which they affect star formation within these molecular clouds.

\end{abstract}


\section{Introduction}\label{sec:intro}
Molecular Clouds (MCs) are the {\rev birthplaces} of stars. These clouds are typically cold (10-30K) and have dense sub-regions within them that may collapse under gravity to form stars. The evolution and efficiency of star formation within MCs are regulated by a number of factors, primarily gravity, turbulence, and magnetic fields \citep{mckee_2007}. Among these physical processes, magnetic fields are perhaps the least well understood, and this is largely because magnetic fields are very difficult to observe directly. 
Although the line-of-sight component of a magnetic field can be measured by Zeeman spectral line splitting \citep{Crutcher}, the width of the line splitting is typically much less than the thermally broadened line width, which makes detection of Zeeman splitting extremely difficult. The long observation times required also make Zeeman observations impractical for producing cloud-scale maps.

As  an alternative, linearly polarized thermal dust emission is commonly used to create large-scale maps to study MC-scale magnetic fields.  This technique relies heavily on the orientation of their dust grains, as dust grains tend to align with their minor axes oriented parallel to the local magnetic field lines.  This alignment is most likely caused by radiative torques from the local radiation field (see \citealt{Andersson2015} for a review). This process creates a net linear polarization orientation of the emitted light that is perpendicular to the magnetic field direction projected on the sky.

By measuring linear polarization of the sub-mm radiation emitted by dust grains within the molecular cloud and rotating the polarization orientation by $90^\circ$, we can map the corresponding magnetic field orientation projected on the plane of sky. {\rev These measured orientation values represent the averaged magnetic field orientation within the volume of the cloud probed by the telescope beam, and are most sensitive to regions of high dust emissivity and efficient grain alignment.}  

In addition to the inferred plane-of-sky magnetic field orientation, there are several other polarization parameters that can be used to study the structure and geometry of magnetic fields in molecular clouds. The polarization fraction ($p$)~of the emission is the fraction of observed radiation that is linearly polarized. $p$~is sensitive to the efficiency with which grains are aligned with respect to the magnetic field, the degree of disorder in the plane-of-sky magnetic field within the dust column probed by the polarimeter, and the  inclination angle of the cloud's magnetic field with respect to the plane of the sky. In addition, the local polarization angle dispersion ($\mathcal{S}$)~is used to measure the disorder in the projected magnetic field orientation at a given location in the cloud \citep{Fissel}. In addition to these polarimetric properties, we also consider the hydrogen column density ($N_{\rm H}$) in our analysis. This quantity is used as a proxy for the total mass surface density of our clouds, and is thus useful for characterising the gas substructure of MCs.

{\rev Lower resolution polarization studies, such as the $1^\circ$ resolution, all-sky analysis of \citet{PlanckXIX}, or the BLASTPol $2'.5$ resolution study of the Vela\,C giant molecular cloud (GMC) \citep{Fissel}, have identified several correlations between \textit{p}, $\mathcal{S}$, and $N_{\rm H}$.} The first trend is a negative correlation between \textit{p} and $\mathcal{S}$:  as $\mathcal{S}$ increases, \textit{p} decreases.
This trend could be related to the differences in the 3D geometry of the magnetic field in different parts of the cloud as $p$~is proportional to $\cos^2\gamma$, where $\gamma$ is the inclination angle of the magnetic field with respect to the plane of the sky \citep{Hildebrand1988}.  For cloud sightlines where the magnetic field is nearly parallel to the line-of-sight, $p$~ values tend to be lower, and upon projection onto the plane-of-sky the angles can vary greatly between adjacent sightlines, leading to large values of $\mathcal{S}$.

In addition, a weak, disordered magnetic field provides little resistance to turbulent motions, and can be easily driven to a highly disordered state, which produces large values for $\mathcal{S}$. This in turn leads to significantly lower \textit{p} values. Strong magnetic fields are able to resist turbulent motions that are perpendicular to their field lines, and thus tend to have lower values for $\mathcal{S}$. 
Low \textit{p} and high $\mathcal{S}$ values can therefore indicate a weak/disordered magnetic field and/or a nearly line-of-sight magnetic field orientation, while high \textit{p} values and low $\mathcal{S}$~values can indicate a strongly ordered magnetic field and/or a nearly plane-of-sky magnetic field orientation.

The second observed trend that we will investigate is the anti-correlation between polarization fraction $p$ and hydrogen column density $N_{\mathrm{H}}$ \citep{PlanckXIX, Fissel}. Dust grains are believed to be aligned with respect to the local magnetic fields by the effect of radiative torques \citep{Andersson2015}. This process may become less efficient in regions of high column density, because photons that can provide these alignment torques are {\rev more likely to be scattered and/or absorbed within higher column density sightlines \citep{King2019}}. This shielding process thus results in lower $p$ values in regions of high $N_{\mathrm{H}}$ \citep{PlanckXX, PlanckXIX}. In \citet{king18}, this correlation could not be explained purely in terms of magnetic field alignment or strength alone, but it was found in \citet{King2019} that introducing grain alignment efficiency may be able to explain it. Correlations between $\mathcal{S}$ and $\mathrm{\textit{N}_H}$ have not to date been firmly established \citep{Fissel}, and so we attempt to contribute to this debate through 2D Kernel Density Estimates involving these two variables.

In \citet{king18}, the authors examined synthetic polarization observations of two magnetized cloud formation simulations generated using the {\tt Athena} MHD code \citep{Chen2016}, with the goal of reproducing the correlations between $p$, $N_\textrm{H}$, and $\mathcal{S}$~found in observations of the Vela\,C GMC in \citet{Fissel}. In order to reproduce the high levels of $\mathcal{S}$, large range of $p$, and the level of anti-correlation between $p$~and $\mathcal{S}$, the authors speculated that Vela\,C must have either a weak magnetic field or a stronger field that is highly inclined with respect to the plane of the sky. Other studies analyzing the orientation of cloud structure with respect to the magnetic field \citep{Soler2017, Fissel19} have suggested that Vela\,C has a reasonably strong cloud-scale magnetic field, and so it seems more likely that Vela\,C's magnetic field must be significantly inclined with respect to the plane of the sky \citep{king18, Chen}. 

Here, we extended the comparisons of the $p$, $N_{\mathrm H}$, and $\mathcal{S}$~distributions to a larger sample size of molecular clouds by including {\em Planck} 353\,GHz polarization observations of eight nearby clouds.
In this way, we will determine if the results of \citet{king18} are consistent with a larger sample size of nearby MCs. The eight \textit{Planck} clouds included in this study are all nearby and relatively low mass, {\rev while Vela\,C is more distant and massive  ($M_\text{Vela}>\,10^5\,M_{\sun}$)}. Significantly higher resolution (2.5\arcmin~as opposed to 15\arcmin~for the {\em Planck} observations) means that Vela\,C has a linear resolution almost equal to the closest of the \textit{Planck} clouds. Linear resolution and other such cloud-specific quantities are listed in Table~\ref{tab:N}.

In this paper, we use our increased sample size to provide more stringent tests of the analyses presented in \citet{king18, King2019, Chen, Fissel19}. We also compare our results to multiple intermediate-results Planck papers, specifically \citet{PlanckXX, PlanckXIX} and \citet{PlanckXII}. 
This paper is organized as follows:~ Section \ref{sec:Obs_dat} - Observations and Data Reduction; ~Section \ref{sec:Polar_Comp} -  Comparison of polarization properties between different clouds; ~Section \ref{sec:Compare_king18} - Comparison with synthetic polarization observations of 3D MHD simulations reported in \cite{king18}; ~Section \ref{sec:Conclusions} - Conclusions. In Appendix \ref{appendix:A}, we discuss how our results are influenced by different methods selecting cloud sightlines. 

\begin{figure*}
    \centering
    \includegraphics[width=0.97\textwidth]{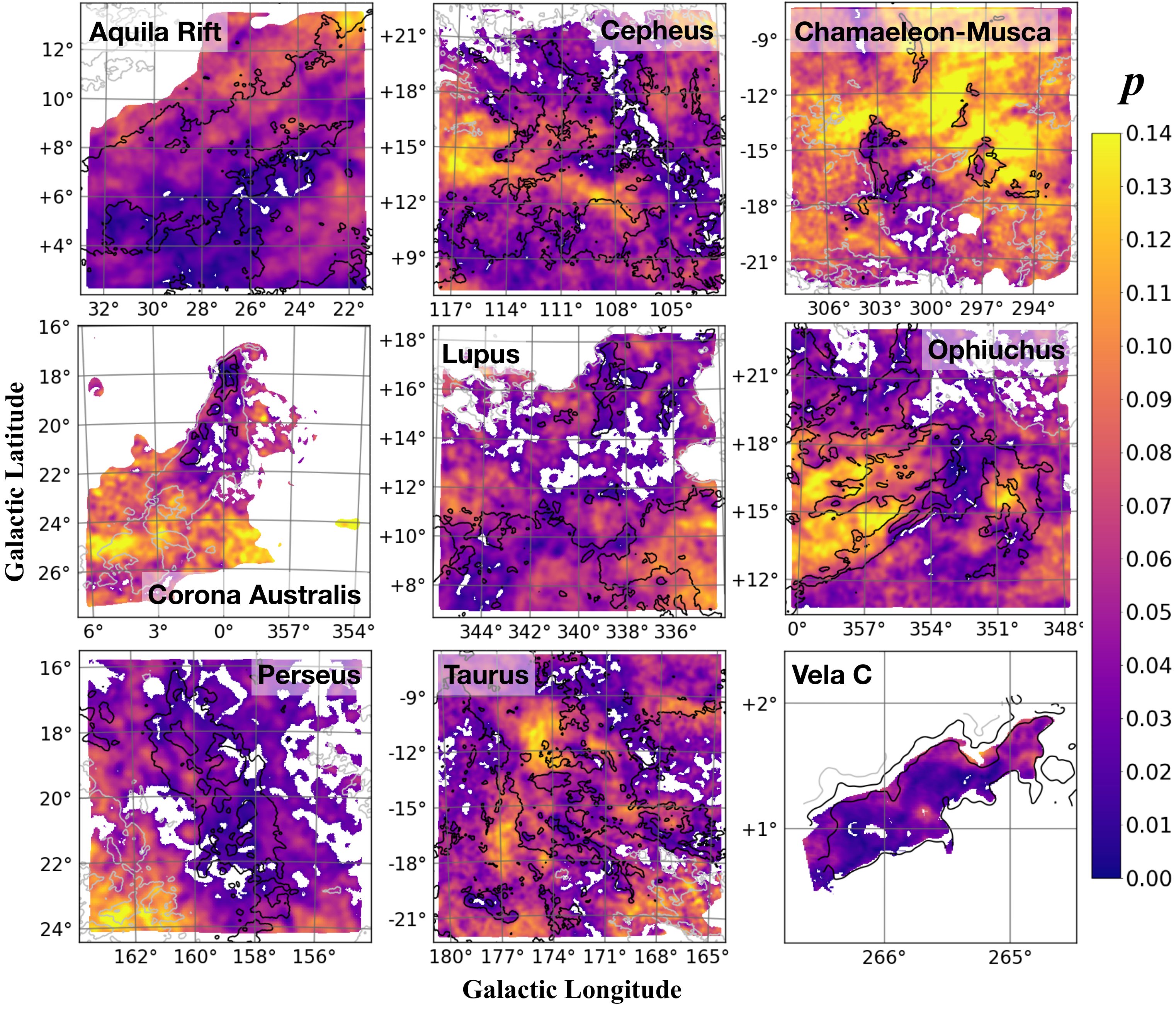}
    \caption{Maps of polarization fraction, as calculated using Equation \ref{eq:p}. Each map has been masked according to the process described in \ref{sec:masking}. Contours are shown for $\mathrm{\textit{N}_H}$. The three contour levels are ${\rev \log}(N_{\rm H}/\rm{cm}^{-2})= 21$ ({\it gray}), $21.5$ ({\it black}), and $22$ ({\it black}). {\rev Throughout this paper, "log" will be used to mean "log base 10" ("log$_{10}$").}}
    \label{fig:p_maps}
\end{figure*}

\begin{figure*}
    \centering
    \includegraphics[width=0.97\textwidth]{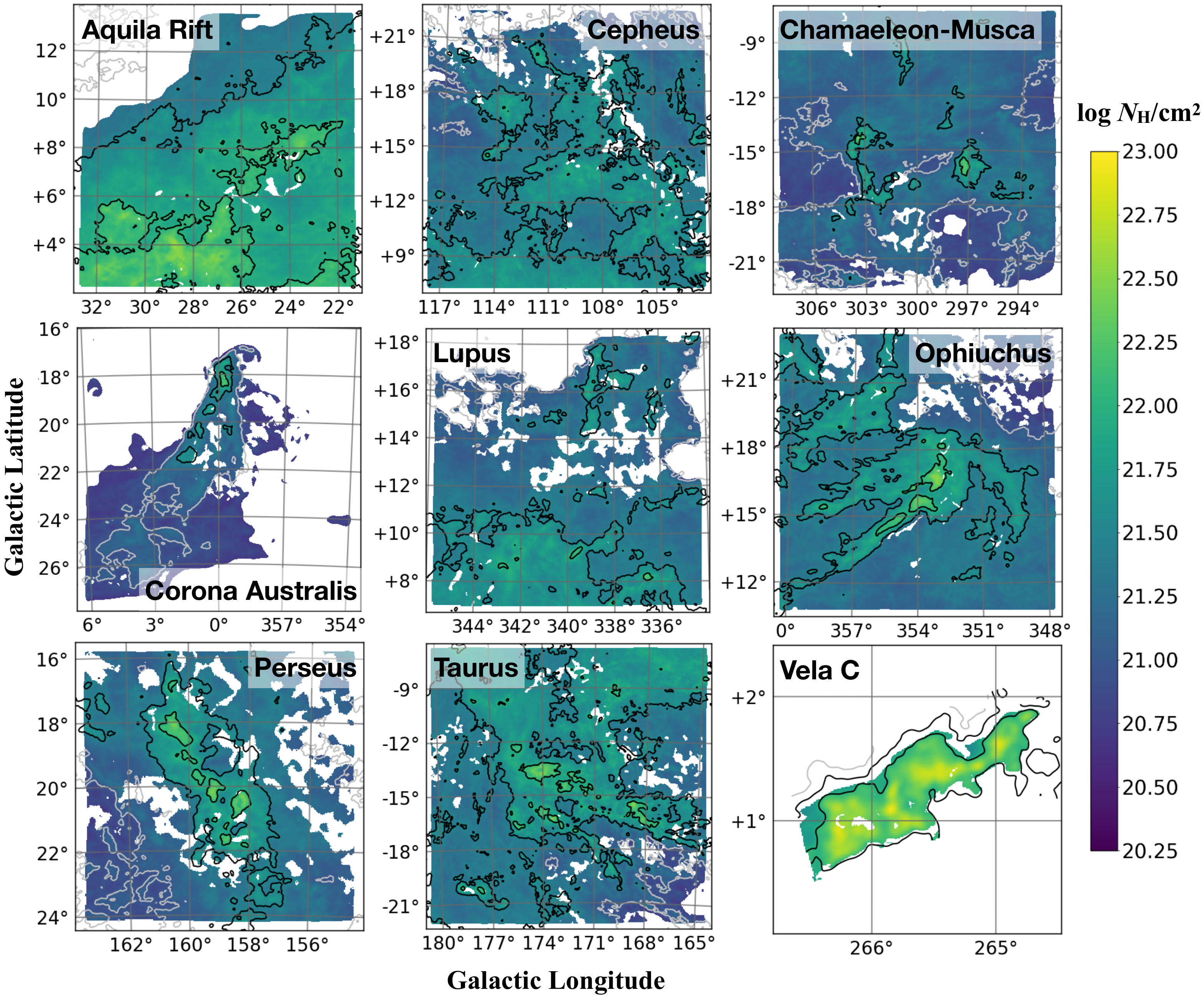}
    \caption{ Maps of hydrogen column density $(\mathrm{\textit{N}_H})$, as obtained from \textit{Planck} and BLASTPol observations, displayed in Galactic coordinates. These $(\mathrm{\textit{N}_H})$ maps from Planck were derived from a 353 GHz optical depth ($\tau_{353}\,GHz$) map using Equation \ref{eq:tau}, while the Vela\,C $(\mathrm{\textit{N}_H})$ map was derived from Herschel SPIRE and PACS submm maps. Each map has been masked according to the process described in Section \ref{sec:masking}, which entailed first smoothing the maps to 15$\arcmin$ FWHM and then removing sightlines which contained values below a specific threshold. Each cloud's unique threshold value is presented in Table \ref{tab:N}. Contours are shown for $\mathrm{\textit{N}_H}$, and the three contour levels are ${\rev \log}(N_{\rm H}/\rm{cm}^{-2})= 21$ ({\it gray}), $21.5$ ({\it black}), and $22$ ({\it black}).}
    \label{fig:N_maps}
\end{figure*}

\begin{figure*}
    \centering
    \includegraphics[width=0.97\textwidth]{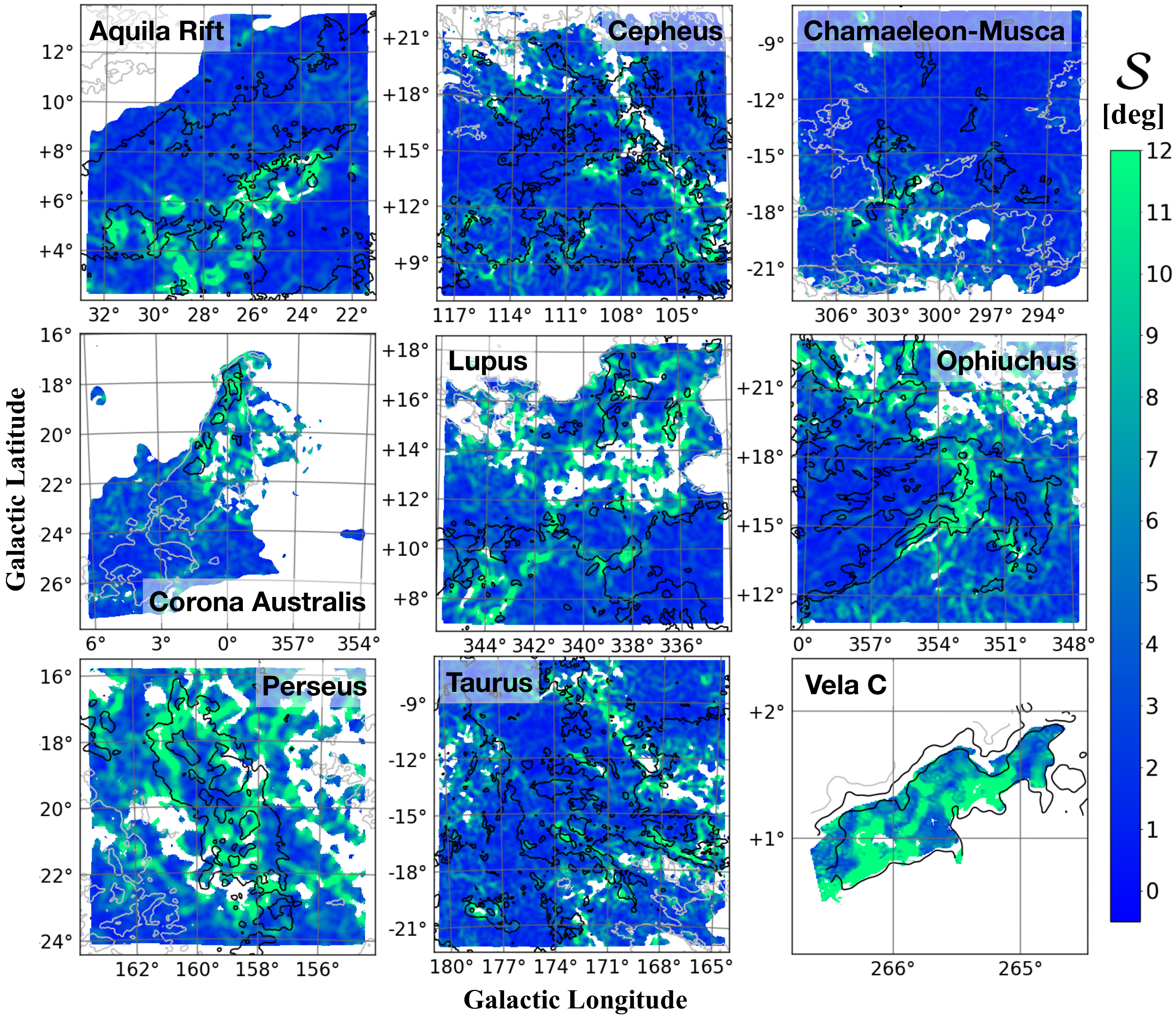}
    \caption{{\rev Maps of dispersion in polarization angles}, as calculated using Equation \ref{eq:S}. Each map has been masked according to the process described in \ref{sec:masking}. Contours are shown for $\mathrm{\textit{N}_H}$. The three contour levels are ${\rev \log} (N_{\rm H}/\rm{cm}^{-2})= 21$ ({\it gray}), $21.5$ ({\it black}), and $22$ ({\it black}).}
    \label{fig:S_maps}
\end{figure*}

\section{Observations \& Data Reduction}\label{sec:Obs_dat}

\begin{table*}
\begin{center}\caption{This table displays the  geometric mean ($\mu_G$), median, and arithmetic standard deviation value ($\sigma$) of each cloud's logarithm of the hydrogen column density $\mathrm{\textit{N}_H}$, as well as the threshold values used to mask cloud regions with low $N_{\rm H}$ as described in Section \ref{sec:masking}. Also listed are the distance to each cloud, the corresponding linear resolution of each observation, and the total observed area for each cloud.  The distance measurements are the median of the Gaia-informed, reddening-based distances  towards different cloud sightlines from \citet{Gaia}, each of which has $\leq~10\%$~errors. Note that some clouds such as the Aquila Rift and Cepheus contain molecular cloud structures at very different distances.}

\label{tab:N}
\begin{tabular}{c c c c c c c c c c c}
\hline 
 & $\mu_G({\rev \log}(N_{\mathrm H}/$ & ${\rev \log}(N_{\mathrm H, {\rm med}} /$ &  $\sigma_{\rev \log} (N_{\mathrm H} /$ &  Threshold $({\rev \log}(N_{\mathrm H} / $ & Distance & Lin. Res. & Obs. Area\\
   &  cm$^{-2}))$  & cm$^{-2})$  &  $_{{\rm cm}^{-2})}$ &  cm$^{-2}))$ & \small[pc] & \small[pc] & \small[pc$^2$] \\
\hline
 AquilaRift       & 21.79 & 21.84 &  0.26 & 21.26 & $ 477$ &  2.1 & 8329\\
 Cepheus         & 21.43 & 21.47 &  0.19 & 20.88 & $ 375$ &  1.6
 & 15910\\
 Chamaeleon-Musca & 21.17 & 21.18 &  0.18 & 20.80 & $ 190$ &  0.8 & 2104\\ 
 Corona Australis & 20.99 & 20.95 &  0.21 & 20.73 & $ 155$ &  0.7
& 1302\\ 
 Lupus            & 21.39 & 21.39 &  0.20 & 20.98 & $ 160$ &  0.7 & 883\\ 
 Ophiuchus        & 21.40 & 21.39 &   0.24 & 20.75 & $ 139$ &  0.6 & 1042\\ 
 Perseus          & 21.33 & 21.32 &  0.27 &20.72 & $ 284$ &  1.2 & 2244\\ 
 Taurus           & 21.49 & 21.48 &  0.24 & 20.79 &  $ 148$ &  0.6 & 1611\\ 
 Vela\,C          & 22.04 & 22.04 &  0.38 & 21.60 & $ 931$ &  0.7 & 1136\\
\hline
\end{tabular}
\end{center}
\end{table*}

In this paper, we analyze thermal dust emission polarization observations of nine nearby MCs.  
For eight of the clouds we use 353\, GHz 
polarization maps from the \textit{Planck} satellite, first presented in \citet{PlanckXXXV}. Our study includes the same set of MCs as \citet{PlanckXXXV} (Aquila Rift, Cepheus, Chamaeleon-Musca, Corona Australis, Lupus, Ophiuchus, Perseus, and Taurus), with the exception of the Orion Molecular Cloud and IC5146. IC5146 was excluded as it is fairly close to the Galactic plane, and 
as such its Stokes $Q$~and $U$ maps had comparable signal to the {\em Planck} polarization maps of diffuse ISM at the same Galactic latitudes. Orion was excluded because it is an evolved, high-mass star forming region, where the magnetic field geometry has likely been strongly affected by feedback from previous high-mass star formation \citep{Soler2018}. We compare these {\em Planck} polarization maps to 500\,$\mu{\rm m}$ polarization maps of the Vela\,C cloud, obtained with the higher-resolution Balloon-borne Large Aperture Submillimeter Telescope for Polarimetry \citep[BLASTPol;][]{BLASTPol2012,Fissel}, which mapped four of the five sub-regions of Vela\,C identified by \citet{Hill}: the South-Nest, South-Ridge, Centre-Nest, and Centre-Ridge. This young GMC has a comparable mass to Orion at $M \sim 10^5M_{\sun}$ \citep{Yamaguchi}, but it is further away \citep[$d\sim 933$~pc;][]{Fissel19} and appears to be much less evolved compared to the Orion Molecular Cloud \citep{Hill}.

The \textit{Planck} data include individual maps of column density of atomic hydrogen $N_{\rm H}$ and the linear Stokes parameters $I$, $Q$, and $U$. The Stokes $I$, $Q$, and $U$~were mapped to a full width at half maximum (FWHM) resolution of 15\arcmin, while the column density maps have a higher resolution of 5\arcmin. The higher resolution $N_{\rm H}$ maps were derived from spectral fits to total intensity maps, which have a much higher SNR than the polarization data. In Table \ref{tab:N}, we list the distance to each cloud and the corresponding linear resolution of each polarization map, which range from 0.6\,pc (Taurus, Lupus, Ophiuchus) to 2.1
\,pc (Aquila Rift).

Using this data, we created maps of polarization fraction (\textit{p}) and dispersion in polarization angles ($\mathcal{S}$). The \textit{p} maps were created using: 
\begin{equation}\label{eq:p}
p = \frac{\sqrt{Q^2 + U^2}}{I}.
\end{equation}
$\mathcal{S}$ was calculated by taking the difference in the polarization angles for all points within a specified lag scale around each pixel, as discussed in Appendix D of \citet{PlanckXXXV}:

\begin{equation}\label{eq:S}
    \mathcal{S}(x,\delta) = \sqrt{\frac{1}{N}\sum_\text{i=1}^N(\Delta\Psi_{x,i})^2},
\end{equation}
where $\delta$~is the lag scale and
\begin{equation}\label{eq:psi}
    \Delta\Psi_{x,i} = \frac{1}{2}\arctan(Q_iU_x - Q_xU_i,\text{  }Q_iQ_x +  U_iU_x)
\end{equation}
is the difference in polarization angle between a given map location $x$ and a nearby map location $i$.
The lag scale we used in our calculations of $\mathcal{S}$ was equal to each observation's resolution: $15\arcmin$ for the {\em Planck} clouds, and $2.5\arcmin$ for Vela\,C. 
The population statistics of $p$~and $\mathcal{S}$~are presented below in Table \ref{tab:p_S}.

The column density map of Vela\,C was derived from dust spectral fits to total intensity maps at 160, 250, 350, and 500 $\mu$m from the {\it Herschel} telescope and smoothed to 2.5\arcmin~FWHM resolution, as described in \citet{Fissel}. {\rev Following the convention in \cite{PlanckXXXV},} the column density maps from \textit{Planck} were derived from a 353 GHz optical depth ($\tau_\text{353}$) map \citep{PlanckXI}, using the relationship: 
\begin{equation}\label{eq:tau}
    \tau_\text{353}/N_\textrm{H} = 1.2 \times 10^{-26}\textrm{cm}^2,
\end{equation}
while $\tau_\text{353}$ was derived from fits to the {\rev 353, 545, and 857 GHz \textit{Planck} and IRAS 100$\mu$m observations} using a modified black body spectrum. {\rev We note that \citet{PlanckXI} finds variations in dust opacity vs.~$N_{\rm H}$, especially between the diffuse and denser phases of the ISM (with the transition at approximately $N_{\rm H}$\,$\approx\,10^{21}$\,cm$^{-2}$), but we note that our applied column density threshold excludes most diffuse ISM sightlines.}

Note that though the BLASTPol polarization maps of Vela\,C have a resolution of 2.5\arcmin, which is higher than the 15\arcmin~resolution of the \textit{Planck} maps,
the linear resolution is comparable to those of the nearer MCs observed with {\em Planck} (see Table \ref{tab:N}) because Vela\,C is much further away. 

\begin{table*}
\begin{center}
\caption{The geometric mean, median, arithmetic and log of geometric standard deviation, for both the \textit{p} and $\mathcal{S}$ polarization parameters. The notation is chosen to match those used in \citet{king18}, Table 7: $\mu_G(p)$ represents the geometric mean of $p$, $p_{\rm med}$ represents the median of $p$, $\sigma_p$ represents the standard deviation of $p$, and $\log\sigma_G(p)$ represents the $\log$ of the geometric standard deviation of $p$. {\rev The same notation format is applied to both $\mathcal{S}$ and $N_{\rm H}$ throughout this paper.}}
\label{tab:p_S}
\begin{tabular}{c c c c c c c c c c c c}
\hline
 & $\mu_G(p)$ & $p_\text{med}$ & $\sigma_p$ & $\log\sigma_G(p)$ & $\mu_G(\mathcal{S})$ & ${\cal S}_\text{med}$ & $\sigma_{\cal S}$ & $\log\sigma_G(\mathcal{S})$ \\
\hline
AquilaRift        & 0.037 & 0.044 & 0.023 & 0.282 & 5.30$^\circ$ & 4.84$^\circ$ & 7.67$^\circ$ & 0.326 \\ 
 
 Cepheus          & 0.046 & 0.047 & 0.026 & 0.223 & 5.62$^\circ$ & 5.43$^\circ$ & 5.86$^\circ$ & 0.290 \\ 
 
 Chamaeleon-Musca & 0.084 & 0.091 & 0.030 & 0.174 & 3.80$^\circ$ & 3.62$^\circ$ & 6.61$^\circ$ & 0.276 \\ 
 
 Corona Australis & 0.069 & 0.075 & 0.030 & 0.208 & 7.19$^\circ$ & 6.90$^\circ$ & 6.61$^\circ$ & 0.270 \\ 
 
 Lupus            & 0.044 & 0.047 & 0.022 & 0.216 & 7.68$^\circ$ & 7.48$^\circ$ & 8.35$^\circ$ & 0.310 \\ 
 
 Ophiuchus        & 0.050 & 0.051 & 0.029 & 0.240 & 7.34$^\circ$ & 7.18$^\circ$ & 8.34$^\circ$ & 0.312 \\ 
 
 Perseus          & 0.039 & 0.038 & 0.027 & 0.248 & 11.13$^\circ$ & 10.93$^\circ$ & 9.43$^\circ$ & 0.280 \\

 Taurus           & 0.048 & 0.050 & 0.024 & 0.212 & 6.54$^\circ$ & 6.29$^\circ$ & 6.27$^\circ$ & 0.276 \\ 

 Vela\,C           & 0.033 & 0.032 & 1.1 & 0.368 & 10.26$^\circ$ & 9.36$^\circ$ & 10.31$^\circ$ & 0.295 \\
\hline
\end{tabular}
\end{center}
\end{table*}

\subsection{Sightline Selection Criteria}\label{sec:masking}
Selecting a limited number of sightlines is important in this analysis, as we are attempting to analyze solely the polarization properties of molecular clouds, unlike other studies which include {\rev diffuse} ISM {\rev sightlines} in their analysis \citep{PlanckXII}. We therefore apply cuts to remove sightlines where the polarization signal is likely to be tracing mostly the more diffuse ISM, or have low signal to noise.

Our  goal is to only select sightlines where the dust emission is likely to be dominated by the cloud rather than foreground or background dust. {\rev While this process is necessary to exclude the diffuse ISM surrounding our target clouds, it also serves to remove sightlines that have low degrees of statistical significance and would therefore necessitate the debiasing of the polarization data. For more details on the ways in which our masking affects our results, see Appendix \ref{appendix:A}.}

We find these {\rev target} regions {\rev by first} comparing the  mean $\mathrm{\textit{N}_H}$ of our {\em Planck} maps to  those of diffuse dust emission. 
\citet{PlanckXXXV} estimated the  contribution of background/foreground dust by  observing a relatively empty area of the sky at the same Galactic latitude as each of their clouds, and assuming that the mean $\mathrm{\textit{N}_H}$ value in this reference region represented the mean $\mathrm{\textit{N}_H}$ of the diffuse ISM around said cloud. We have used the same reference maps in our analysis to identify threshold column density levels for each cloud, which are listed in Table \ref{tab:N}. We  compare the $\mathrm{\textit{N}_H}$ of each cloud to the threshold value derived from its corresponding reference region:  any pixel in a cloud's column density map (smoothed to 15\arcmin~FWHM) that was below this cutoff value was masked, and therefore excluded from future analysis.\footnote{Note that the one-dimensional column density PDFs shown in Figure \ref{fig:1D} do not include any sightline cuts.  This was done in order to show  the distribution of the cloud column density unbiased by the cloud polarization levels. The background $\mathrm{\textit{N}_H}$~threshold levels for each cloud are  presented in Table \ref{tab:N} and indicated in the top row of Figure \ref{fig:1D} with dashed vertical lines.}

In addition, we applied  polarization-based cuts to ensure that  the sightlines we were evaluating have statistically significant polarization detections.  We compared the strength of each sightline's polarized intensity ($P$) to its uncertainty and masked out any sightlines for which:

\begin{equation}\label{eq:PI}
\dfrac{P}{\sigma_{P}} < 3,
\end{equation}

where $\sigma_{P}$ is the uncertainty in $P$. Figure \ref{fig:p_maps} shows the $p$~maps for the dust sightlines that pass all of our selection criteria. 
  
Because Vela\,C is near the Galactic plane, the values included in this study also only include the regions within the dense cloud sub-regions defined by \citet{Hill}, where the contribution from the diffuse ISM along the same sightlines is not significant.

We note that our sightline selection will tend to bias our observations {\rev towards regions of higher polarization fraction. The cuts based on column density eliminate low-column dust sightlines where the polarization fraction tends to be high.  However, the cuts based on polarized intensity tend to eliminate a larger fraction of sightlines with low $p$ and high $\mathcal{S}$ in intermediate and high column density regions.} 
We discuss in detail how different choices of sightline selection method affect our results in Appendix \ref{appendix:A}.

\vspace{0.5cm}

\section{Comparison of Polarization Properties}\label{sec:Polar_Comp}

In this section, we present our analysis of the {\em Planck} polarimetric data. Note that though we are investigating regions that have been previously studied in \citet{PlanckXIX, PlanckXX} and  \citet{PlanckXII}, the {\em Planck} papers aimed to characterize the magnetic fields in molecular gas clouds while also including sightlines that probe the diffuse interstellar medium (ISM).
In contrast,
we have attempted to only select sightlines where the dust emission is associated with these particular molecular clouds.  
Furthermore, our sightline masking strategy described in Section \ref{sec:masking} predominantly removes sightlines with low polarization fraction ($p$). By masking out low polarization regions in an attempt to reduce the contributions of background noise, we correspondingly mask out the regions of highest dispersion in polarization angles ($\mathcal{S}$). This masking process thus leads to us having significantly lower $\mathcal{S}$ values than those recorded in the {\em Planck} papers.  The effects of  sightline selection choices on our results are discussed in more detail in Appendix \ref{appendix:A}.

We also note that there is a difference in resolution and lag scale ($\delta$) between our work and previous studies. \citet{PlanckXIX} and \citet{PlanckXX} state that $\mathcal{S}$ increases with $\delta$. Thus, the resolution and $\delta$ used in \citet{PlanckXIX}, a resolution of $1^\circ~{\rm and}~\delta = 30\arcmin$, are likely to produce higher $\mathcal{S}$ values than our study, which used resolution values of $15\arcmin~(Planck)$ and $2.5\arcmin$ (BLASTPol) for their corresponding lag scales. This effect likely decreased our $\mathcal{S}$ values compared to \citet{PlanckXIX}, but \citet{PlanckXX} uses the same resolution of $15\arcmin$ as ours, and a very comparable lag scale of $\delta = 16\arcmin$. This means that the difference in $\mathcal{S}$ values between our study and those of \citet{PlanckXX} will be almost entirely due to the differences in masking.

\subsection{Probability Distributions of Polarimetric Observables}
\label{sec:1Dplot}

We first compare the properties of each cloud by considering the distribution of our three observable values, $\mathrm{\textit{N}_H}$, \textit{p}, and $\mathcal{S}$, individually. These plots were created using a Gaussian KDE from the \texttt{astropy.convolution} package in Python \citep{astropy:2013, astropy:2018}, which takes an input variable and calculates its probability density function (PDF). These PDFs provide the relative probability density, which we refer to as $f(x)$, at all values, which is a measure of how likely a random data point is to fall within the given range of $x$ values. Figure \ref{fig:1D} shows the PDFs for $\mathrm{\textit{N}_H}$, \textit{p}, and $\mathcal{S}$ on a logarithmic scale. 

The PDFs are useful as they provide a basic characterization of the population of each observable's distribution, including its width and peak (most probable value) within a given cloud. The vertical dashed lines in the top row of Figure \ref{fig:1D}, the log($\mathrm{\textit{N}_H}$) PDFs, indicate each cloud's respective background $\mathrm{\textit{N}_H}$ cutoff value, which were obtained from Appendix B of \citet{PlanckXXXV},  are listed in Table \ref{tab:N}, and represent our estimates of the  average column density contribution from foreground and background dust.
In addition to these plots, the median, geometric mean $\mu_G()$, arithmetic standard deviation $\sigma$, and the geometric standard deviation {\rev are presented for $\mathrm{\textit{N}_H}$ in Table \ref{tab:N}, and for \textit{p} and $\mathcal{S}$ in Table \ref{tab:p_S}.}  

\subsubsection{PDF of $N_{\rm H}$}

The PDFs  in the top row of Figure \ref{fig:1D} show that there are clear differences in column density distribution among our targeted clouds. On the one hand, Corona Australis' and Chamaeleon-Musca's column densities are relatively low. In the case of Corona Australis,  a large portion of the map's sightlines have $\mathrm{\textit{N}_H}$ values that fall below the background reference value.
On the other hand, Vela\,C, a young GMC, has a significantly higher $N_{\rm H}$ distribution and $\mu_G({\rev \log}(\mathrm{\textit{N}_H}))$ value than any other cloud. The Aquila Rift also has a particularly high $\mu_G({\rev \log}(N_{\rm H}))$ value of 21.79, a value that is significantly above  those of the other {\em Planck} clouds such as Cepheus, Ophiuchus, Lupus, Taurus, and Perseus  (as can be seen in Table \ref{tab:N}). It should once again be noted that the $\rm{\textit{N}_H}$ PDFs were calculated before masking, and after being smoothed to a 15$\arcmin$ FWHM resolution. These same trends can also be observed in Fig. \ref{fig:N_maps}, wherein each cloud's $\rm{\textit{N}_H}$ map is presented with contours to represent ${\rev \log} (N_{\rm H}/\rm{cm}^{-2})= 21$ ({\it gray}), $21.5$ ({\it black}), and $22$ ({\it black}).

\subsubsection{PDF of $p$}
\label{sect:angles}

The polarization fraction PDFs show that Chamaeleon-Musca has particularly high polarization levels compared to the other clouds. Of our cloud sample, Vela\,C, Perseus, and the Aquila Rift show the lowest $p$ value distributions (see Table \ref{tab:p_S}). As discussed in Section \ref{sec:intro}, clouds with magnetic fields that are strong compared to turbulent gas motions will tend to have a high polarization fraction if the mean ordered magnetic field is not significantly inclined \citep{king18}. However, there is a degeneracy with the viewing angle as the polarization fraction is roughly proportional to $\cos^2(\gamma)$, where $\gamma$~is the averaged inclination angle between the magnetic field and the plane of the sky. Viewing the cloud along a sightline nearly parallel to the magnetic field will tend to result in very low measured polarization levels. 
It is likely that among our selected clouds there would be a range of viewing geometries and magnetic field strengths, resulting in a range of polarization fraction measurements.

\begin{table}
\begin{center}
\caption{Inclination angles of the cloud-scale magnetic fields derived from the distribution of polarization fraction, as described in \citet{Chen}.}
\label{tab:pDA}
\begin{tabular}{c c c c c}
\hline
 & $p_{\mathrm{max}}$ & $\gamma_{\mathrm{obs}}^{\wedge}$ & ${\gamma^{\wedge}_{\mathrm{obs},S<\langle S\rangle}}$ & estimated $\gamma_{\overline{\mathbf{B}}}^{\dagger}$ \\
\hline
AquilaRift & 0.15 & 52.4$^\circ$ & 51.2$^\circ$ & 53$^\circ$ \\
Cepheus & 0.16 & 56.5$^\circ$ & 50.0$^\circ$ & 51$^\circ$ \\
Chamaeleon-Musca & 0.19 & 41.0$^\circ$ & 35.1$^\circ$ & \ 17$^\circ$ \\
Corona Australis & 0.19 & 50.4$^\circ$ & 44.3$^\circ$ & \ 38$^\circ$ \\
Lupus & 0.15 & 55.1$^\circ$ & 48.9$^\circ$ & 48$^\circ$ \\
Ophiuchus & 0.16 & 58.3$^\circ$ & 46.8$^\circ$ & 43$^\circ$ \\
Perseus & 0.18 & 62.8$^\circ$ & 57.7$^\circ$ & 68$^\circ$ \\
Taurus & 0.15 & 52.0$^\circ$ & 48.7$^\circ$ & 48$^\circ$ \\
Vela\,C & 0.14 & 63.9$^\circ$ & 52.1$^\circ$ & 56$^\circ$ \\
\hline

\label{tab:inc}
\end{tabular}
\end{center}
\footnotesize{$^\dagger$ As described in \cite{Chen}, there are intrinsic differences between the observed inclination angle $\gamma_\mathrm{obs}$ derived from $p$ and the actual inclination angle $\gamma_{\overline{\mathbf{B}}}$ in the 3D space due to projection effects (see e.g.,~their Figures 5 \& 11). Here we adopted the correlation between $\gamma_{\mathrm{obs}}^{\wedge}$ and $\gamma_{\overline{\mathbf{B}}}$ shown in Figure~11 of \cite{Chen} to get the final estimate of the 3D inclination angle of the cloud-scale magnetic fields for the Planck clouds.}
\end{table}

\begin{center}
\begin{figure*}
\centering
\includegraphics[scale=0.35]{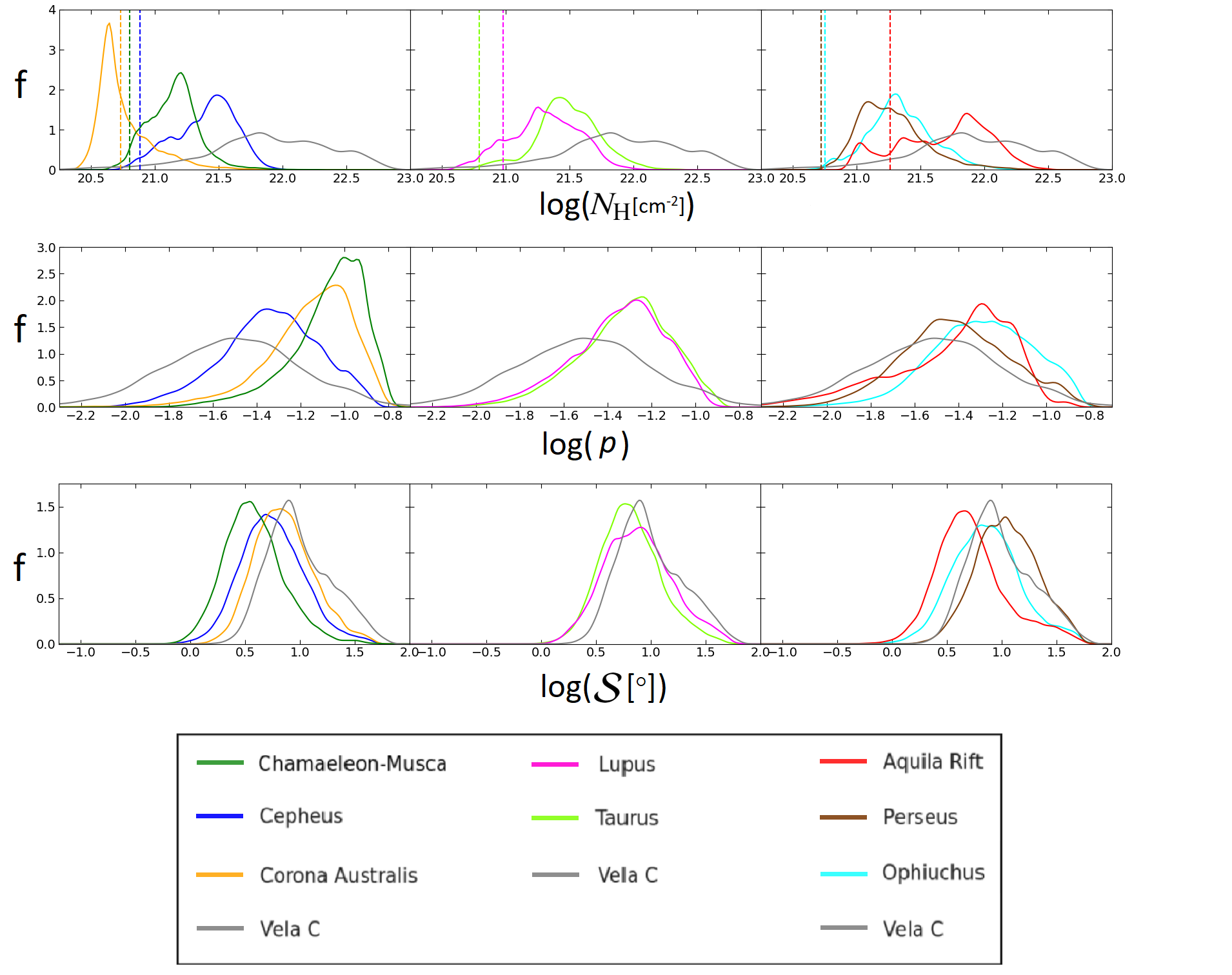}
\caption{PDFs for each cloud's $\mathrm{\textit{N}_H}$ ({\it top}), \textit{p} ({\it middle}), and $\mathcal{S}$ ({\it bottom}) values on the smoothed scale of 15$\arcmin$ FWHM.  The y axis in this figure shows the probability density, labelled here as f. Probability density is defined by the area under the curve which it creates: the probability of an x value being lower than a given quantity within the bounds of the curve is equal to the area under the curve to the left of that quantity. Each column displays the data from a specific subset of clouds,  divided to increase the clarity of the plots. Vela\,C is shown in every plot as a reference for comparison. Note that the vertical dashed lines in the $\mathrm{\textit{N}_H}$ PDFs  indicate the column density threshold that each clouds sightline must be greater than to be included in our analysis, as  displayed in Table \ref{tab:N} and discussed in Section \ref{sec:masking}. The column density $\mathrm{\textit{N}_H}$ PDFs thus include all of each map's sightlines, however only sightlines that passed the selection criteria described in Section \ref{sec:masking} were used for the \textit{p} and $\mathcal{S}$ PDFs  shown in this figure.}
\label{fig:1D}
\end{figure*}
\end{center}

 The distribution of the polarization fraction $p$ within a molecular cloud can also be used to estimate the average inclination angle of the magnetic field in the cloud, as described in \cite{Chen}.  For each of our Planck clouds, we first determined the maximum polarization fraction $p_\mathrm{max}$ within the cloud by examining the 1D PDF of $p$ (see e.g.,~Figure~\ref{fig:1D}). 
{\rev Assuming this value corresponds to sightlines with uniform magnetic field (i.e.,~position angle $\psi={\rm constant}$) completely on the plane of sky (i.e.,~inclination angle $\gamma=0$) along the line of sight and considering the widely adopted dust polarization equations \citep[see e.g.,][]{FiegePudritz2000}:}
\begin{align}
q &= \int n \cos 2\psi \cos^2 \gamma ~dz ,\ \ \ \ 
u = \int n \sin 2\psi \cos^2 \gamma ~dz,\notag\\
p & = p_0 \frac{\sqrt{q^2 + u^2}}{N - p_0 N_2}, \
\ \ \ N_2 = \int n\left(\cos^2 \gamma - \frac{2}{3}\right) ~dz,
\end{align}
one has
\begin{equation}
    p_{\rm max} = \frac{p_0\cos^2\gamma}{1 - p_0\left(\cos^2\gamma-\frac{2}{3}\right)} \Bigg|_{\cos^2\gamma=1} = \frac{p_0}{1 - \frac{1}{3}p_0}.
\end{equation}
We then used equation~(10) of \cite{Chen} to calculate the inclination angle $\gamma_\mathrm{obs}$ using the observed polarization fraction at each pixel, $p_\mathrm{obs}$:
\begin{equation}
    \cos^2\gamma_\mathrm{obs} = \frac{p_\mathrm{obs}\left(1+\frac{2}{3}p_0\right)}{p_0 \left(1+p_\mathrm{obs}\right)}.
\end{equation}
Following the methods of \cite{Chen}, we consider the most probable value of $\gamma_{\mathrm{obs}}$ for all detections of $p_{\mathrm{obs}}$ among the entire cloud as the cloud-scale inclination angle of the magnetic field, $\gamma_{\mathrm{obs}}^{\wedge}$. We also adopted the $\mathcal{S}$-correction proposed in \cite{Chen} to include only regions that are less perturbed (with ${\cal S} < \langle {\cal S}\rangle$, the median value of $\mathcal{S}$), which tend to have smaller errors in $p$-derived inclination angle. The results are listed in Table~\ref{tab:pDA}, including the final estimate of the cloud-scale magnetic field inclination angle $\gamma_{\overline{\mathbf{B}}}$ after considering the errors between the projected $\gamma$ and the actual one in 3D (see the footnote of Table~\ref{tab:pDA}).
{\rev We also note that, while there is no doubt that the value of $p_{\rm max}$ is  influenced by the resolution of the polarization measurement (because larger telescope beams tend to remove extreme values of \textit{p}; see e.g.,~\citealt{king18}), as discussed in \cite{Chen}, the uncertainty in determining $p_{\rm max}$ is unlikely to introduce large deviations in the derived inclination angle, and the projection effect from $\gamma_{\overline{\mathbf{B}}}$ to $\gamma_{\rm obs}$ is relatively more significant.}
 Because of this intrinsic error associated with projection from 3D to 2D, the expected accuracy of this method is only $\sim 10-30\deg$ (see discussions in \citealt{Chen}). Nevertheless, these values provide important input for our discussion on the properties of individual clouds in the next section.

\subsubsection{PDF of $\mathcal{S}$}

As for the $\mathcal{S}$ PDFs, we note that  Chamaeleon-Musca and Perseus peak at the lowest and highest values of ${\cal S}$, with $\mu_G(\mathcal{S}) = 3.8^\circ$ and $11.13^\circ$, respectively. 
In contrast, Chamaeleon-Musca has the highest values of $p$, and the $\mu_G(p)$ of Perseus is the second lowest, only slightly higher than that of Vela C. 

These results are consistent with previous observations of a negative correlation between \textit{p} and $\mathcal{S}$ within individual clouds \citep{Fissel, PlanckXII, king18}. Other clouds also  display a negative correlation between $\mathcal{S}$ and \textit{p} distributions, but  Chamaeleon-Musca and Perseus in particular display the strongest contrast. It is possible that these two clouds show such strong contrasts between $p$~and $\mathcal{S}$ because they may be individual realizations of near-limiting cases: a large magnetic field strength and/or a low inclination angle in the case of Chamaeleon-Musca, and a weak field and/or a high inclination angle in the case of Perseus. Vela\,C is an outlier in the column density PDF, but it also has the lowest median \textit{p} value, along with the second highest median $\mathcal{S}$ value among our sampled clouds. \citet{king18} argued that this combination implies that Vela\,C has an unusually high inclination angle $\gamma$ or potentially a weak magnetic field. These same trends can also be observed in Fig. \ref{fig:S_maps}, wherein each cloud's $\mathcal{S}$ map is presented with contours to represent ${\rev \log} (N_{\rm H}/\rm{cm}^{-2})= 21$ ({\it gray}), $21.5$ ({\it black}), and $22$ ({\it black}).

\subsection{Joint Correlations from 2D Kernel Density Estimates}

\label{sec:Continuous Plots}

\begin{figure*}
\begin{center}
\includegraphics[width=\textwidth]{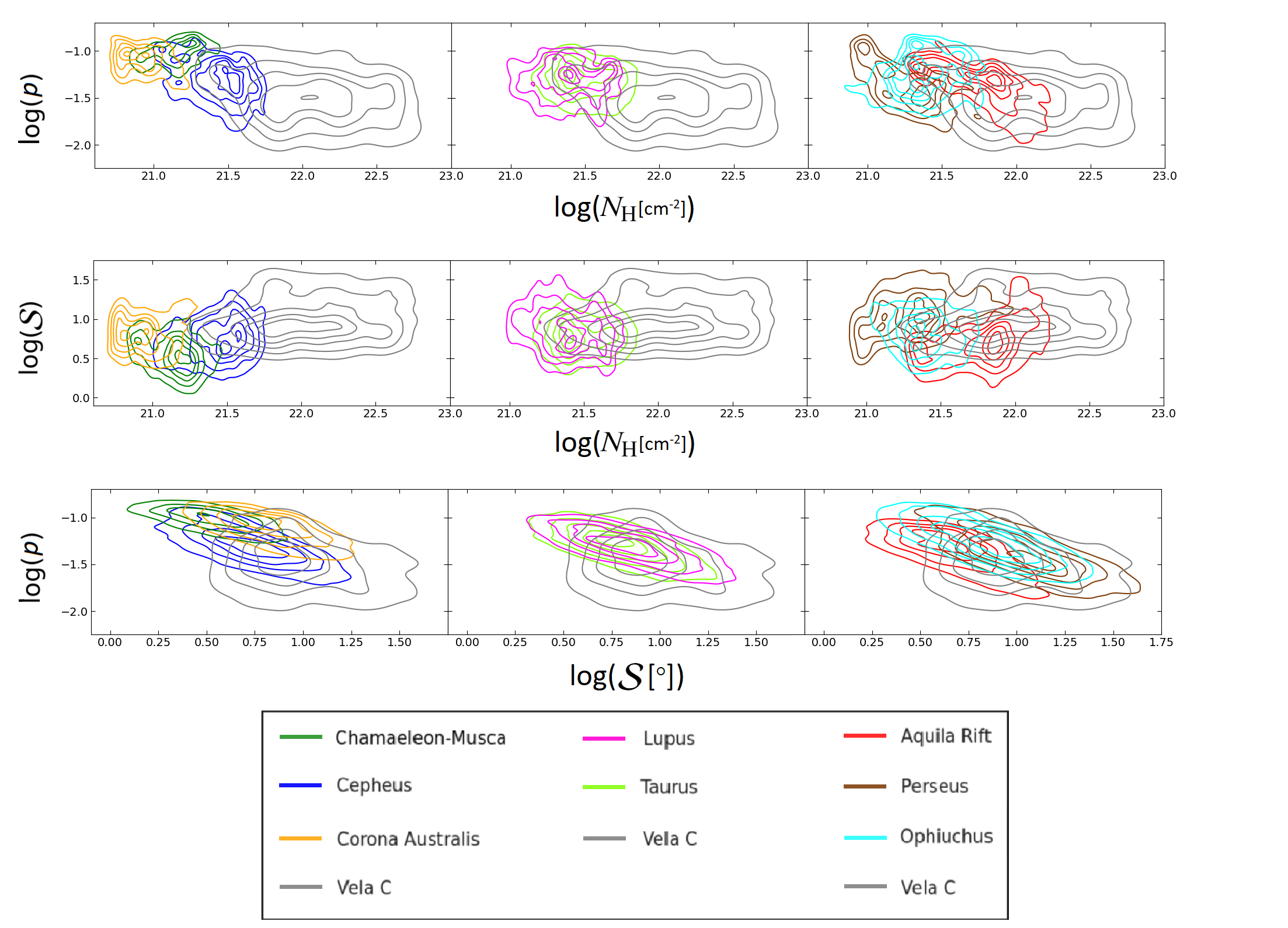}
\caption{Joint PDFs of our polarization observables: {\em top row:} polarization fraction ($p$) vs hydrogen column density ($N_{\rm H}$), {\em middle row:} polarization angle dispersion ($\mathcal{S}$) vs $N_{\rm H}$, {\em bottom row:} $p$ vs $\mathcal{S}$. Each contour colour represent a different molecular cloud, while the gray contours represent the data of the Vela\,C cloud, obtained from \citet{king18} and provided in every plot as a point of comparison.}
\label{fig:2D}
\end{center}
\end{figure*}

In this section, we examine the joint correlations between the polarization observables using KDE as described in \citet{king18}. For these comparisons, the {\em Planck} column density maps have been smoothed to the same 15\arcmin~FWHM resolution as the $p$~and $\mathcal{S}$~maps. The displayed slope values are calculated using the second eigenvector's $x$ and $y$ components, and the eigenvectors themselves are calculated using \texttt{numpy}'s linear algebra package in \texttt{Python}. These slopes thus represent the  correlation between the different pairings of cloud observables.  

The results are presented in Figure~\ref{fig:2D}, with the fitted parameters listed in Table~\ref{tab:Pear_Spear}.

\begin{table*}
\begin{center}
\caption{This table shows the Pearson and Spearman coefficients for the correlations between each of the three sets of polarization comparisons: $p$-$N_{\rm H}$, $\mathcal{S}$-$N_{\rm H}$, and $p$-$\mathcal{S}$. The coefficients are denoted by P and an S subscripts, for Pearson and Spearman respectively.}
\label{tab:Pear_Spear}
\begin{tabular}{c c c c c c c c c c}
\hline
 & $\rho_{\text P, p-N}$ & $\rho_{\text S, p-N}$ & $p-N$ & $\rho_{\text P, \mathcal{S}-N}$ & $\rho_{\text S, \mathcal{S}-N}$ & $\mathcal{S}-N$ & $\rho_{\text P,p-\mathcal{S}}$ & $\rho_{\text S,p-\mathcal{S}}$ & $p-\mathcal{S}$ \\
 & & & Index & & & Index & & & Index \\
\hline
 AquilaRift       & -0.670 & -0.722 & -1.12 & 0.452 & 0.459 & 1.60 & -0.799 & -0.732 & -0.836 \\
 Cepheus          & -0.342 & -0.311 & -0.702  & 0.202 & 0.193 & 4.40 & -0.734 & -0.718 & -0.702 \\
 Chamaeleon-Musca & -0.005 &  0.182 & -0.533 & -0.071 & 0.139 & -12.6 & -0.714 & -0.673 & -0.533 \\
 Corona Australis & -0.627 & -0.453 & -0.687  & 0.352 & 0.212 & 2.00  & -0.685 & -0.670 & -0.687 \\
 Lupus            & -0.153 & -0.123 & -1.63  & 0.120 & 0.144 & -7.69 & -0.746 & -0.742 & -0.619 \\
 Ophiuchus        & -0.077 & -0.027 & -1.23  & 0.022 & -0.038 & 25.9 & -0.762 & -0.763 & -0.708 \\
 Perseus          & -0.594 & -0.598 & -0.882 & 0.272 & 0.285 & 1.20 & -0.718 & -0.719 & -0.846 \\
 Taurus           & -0.153 & -0.156 & -0.460 & 0.050 & -0.039 & 5.50  & -0.695 & -0.676 & -0.688 \\
 Vela\,C           & -0.068 & -0.055 & -1.58 & 0.027 & 0.024 & -3.89 & -0.244 & -0.249 & -0.930 \\
\hline
\end{tabular}
\end{center}
\end{table*}

\subsubsection{The $p-{\cal S}$ correlation}\label{sec:pS}
As previous studies have noted \citep[see e.g.,][]{Fissel, PlanckXX, PlanckXII, PlanckXIX},
we find a negative correlation between log($p$) and log($\mathcal{S}$) for all of our target clouds (bottom row of Figure \ref{fig:2D}). The $p$ vs $\mathcal{S}$ trend was first presented in \citet{PlanckXIX}, and was based off of the distribution of $p$ and $\mathcal{S}$ values from the entire sky. \citet{PlanckXX}'s found a similar correlation for higher resolution maps with observations restricted to 12$^\circ$x12$^\circ$ fields of nearby clouds and diffuse ISM regions.

In comparing the \textit{p} vs $\mathcal{S}$ trends among the various clouds, Vela\,C shows the steepest decrease in $p$~with $\mathcal{S}$~compared to the other clouds (bottom row of Figure \ref{fig:2D}).  All of the clouds  in our study have a negative  $p$ vs $\mathcal{S}$ slope, with a mean value of roughly -0.694, while Vela\,C's $p$-$\mathcal{S}$ correlation has a slope of -0.930, though this may be partially caused by the narrower range of $\mathcal{S}$ in the Vela\,C polarization data. Studies of synthetic polarization observations by \citet{king18} found that  the inclination angle between the line-of-sight and mean magnetic field orientation of  a cloud has a significant effect on the slope of  its \textit{p} vs $\mathcal{S}$ relationship. \citet{king18}, and later \cite{Chen}, propose that the large $p$~vs $\mathcal{S}$~slope of Vela\,C is caused by a large inclination angle of the mean magnetic field orientation with respect to the plane of the sky ($\sim\,60^\circ$).  This is consistent with our findings in Section \ref{sect:angles} where we used the method proposed in \cite{Chen} to estimate the inclination angle of the magnetic field for our clouds, and found that Vela\,C had an estimated mean cloud-scale magnetic field inclination angle of 56$^{\circ}$.

Both \citet{PlanckXX} and \citet{PlanckXIX} argue that for large regions of the ISM where a wide range of magnetic field orientations are seen, the $p$ vs $\mathcal{S}$ slope should be constant.  While we agree that a negative correlation between $p$ and $\mathcal{S}$ is certainly present in all of our target clouds, we find considerable variation in their $p$ vs $\mathcal{S}$~slope values throughout our analysis, which appears to be significantly affected by the inclination angle of each cloud's magnetic field with respect to the plane of sky. We note that this variation in $p$ vs $\cal{S}$ slopes seems to be correlated with inclination angles, as we have estimated lower inclination angles for the clouds with shallower slopes, and higher inclination angles for those that have steeper slopes (see Section \ref{sect:angles} and Section \ref{sec:Compare_king18}).

The strength of this relationship may also be impacted by the sightline selection criteria that we have chosen to employ in Section \ref{sec:masking}. Whereas the \textit{Planck} papers included polarization values from all sightlines, we have applied masks to our data to remove sightlines  with low signal to noise polarization detections and where  most of the emission is likely dominated by  fore- or background diffuse dust. In doing so, however, we may somewhat bias our analysis toward regions of higher polarization. which corresponds to regions of higher $p$ and lower $\cal{S}$.   {\rev As shown in Appendix \ref{appendix:A} and Figure \ref{fig:pS_masking}, applying cuts to eliminate sightlines below the threshold column density tends to remove sightlines with high values of $p$~and $\mathcal{S}$, while applying cuts based on $P$ removes  a larger number of low-$p$, high-$\mathcal{S}$ sightlines.  Applying both selection criteria has the net effect of} reducing the slope of the $p$~vs $\mathcal{S}$~relation {\rev (see Figure \ref{fig:pS_vs_Smasking})}, but the relative ordering of the $p$-$\mathcal{S}$~slope indices remain largely unchanged (e.g.~Chamaeleon-Musca/Perseus always have the shallowest/steepest slope regardless of which sightline selection criteria are used).

\subsubsection{The $p-N_{\rm H}$ correlation}\label{sec:pN}

The column density - polarization fraction correlation has been reported in previous works to be robustly anti-correlated in a variety of targets \citep{Fissel, king18, PlanckXIX}. In our sample, we find  three clouds that do not follow this trend: Chamaeleon-Musca,  Lupus and Taurus (see the top row of Figure \ref{fig:2D}). These clouds have very low statistical correlation between $p$ and $N_\textrm{H}$ (see Table~\ref{tab:Pear_Spear}).
We note that for Chamaeleon-Musca there is very little range in polarization fraction, and the $p$~values are systematically higher than those of other clouds.

The decreasing trend between $p$ and $N_\mathrm{H}$ is usually explained by two processes: de-polarization due to changes in grain alignment efficiency or grain properties and the tangling of magnetic fields. As the $N_\mathrm{H}$ increases, an increasing number of dust grains that become shielded by other dust grains from the photons that have a short enough wavelength to exert radiative alignment torques. This means that \textit{p} values may drop towards the high-$N_\mathrm{H}$ regions, as there are fewer dust grains that are aligned with respect to the local magnetic field and therefore the net polarization measured from the dust column is lower.

The correlation between polarization fraction $p$ and hydrogen column density $N_{\rm H}$ has been studied in many previous polarization studies \citep[e.g.,][]{PlanckXIX, PlanckXX, PlanckXII,king18}.
Among them, \citet{king18} and \citet{PlanckXX} offer support in favor of two different interpretations of this anticorrelation. On the one hand, \citet{PlanckXX} argue that the whole sky trend is reproducible using synthetic observations assuming no variations in the efficiency of dust grain alignment with respect to the magnetic field. This implies that the observed depolarization is caused by magnetic field tangling within the telescope beam; i.e.,~the 3D geometry of the magnetic field. This appears to be consistent with the leading theory of grain alignment, radiative torque alignment \citep[RAT;][]{Cho_Lazarian, Andersson2015}, which demonstrated that large dust grains, specifically those found in molecular clouds, can still be magnetically aligned in regions of high column density. 

On the other hand, \citet{king18} argued that the correlations obtained from synthetic observations of their MHD simulations, which did not include a loss of grain alignment efficiency towards high column density sightlines, could not reproduce the decrease in $p$ observed with increasing $N_{\rm H}$ in Vela\,C. 
The efficiency with which grains are aligned nevertheless depends on the specific grain population in question, the microphysics of grain alignment, and properties of the local radiation field. 
\citet{king18} argued that the lack of agreement between the $p$ vs $N_\textrm{H}$ trends found in Vela\,C and their simulations is primarily due to their assumption of homogeneous grain alignment. In \citet{King2019}, it was further shown that by including a simple analytic model for the decrease in grain alignment efficiency with density, it is possible to reproduce the $p$ vs $N_{\rm H}$ trends observed in the BLASTPol observations of Vela\,C.

There are two major differences between the work in \citet{king18} and \textit{Planck} studies which may contribute to the discrepancies. First, whereas \citet{king18} evaluated the $p-N_{\rm H}$ slope using the entire $p$ value distribution, \citet{PlanckXX} fitted the slope considering only the upper envelope of their p distributions, $p_{\rm max}$, to minimize the statistical impact of sightlines where the magnetic field orientation is significantly inclined from the plane of the sky, which decrease the observed polarization fraction. This is opposed to the fits of log($p$) vs log($N_{\rm H}$) discussed in both this work and \citet{king18}. 

In addition, as mentioned earlier, there is a significant difference in the range of column densities of the regions studied by \citet{PlanckXX} and our work. Most of the sightlines included in \citet{PlanckXX} trace the more diffuse component of the ISM, whereas our study includes only sightlines above a certain column density threshold, and are therefore more likely to include regions where radiative alignment torques are less efficient. Still, the discrepancy between the synthetic observations by \citet{king18} and those presented in \citet{PlanckXX} warrants further study, and could point to a difference in the underlying physics of the simulation, e.g.,~the driving mechanism of turbulence. This discrepancy has important implications for interpreting polarization data. If the decrease in $p$ vs $N_{\rm H}$ is only due to changes in dust grain alignment efficiency, the $p$ vs $N_{\rm H}$ trend can be used to directly probe the dust physics. However, if the $p$ vs $N_{\rm H}$ trend is also affected by the structure of the magnetic field, then it will be more difficult to model the grain alignment efficiency as a function of density.

\subsubsection{The ${\cal S}-N_{\rm H}$ correlation}\label{sec:SN}

Based on the appearance of the $\mathcal{S}$ vs $\mathrm{\textit{N}_H}$ KDE estimates, only a few clouds seem to show a weak positive correlation between $\mathrm{\textit{N}_H}$ and $\mathcal{S}$ (see the middle row of Figure~\ref{fig:2D}). We therefore considered the Pearson and Spearman coefficients here, which are measurements of the level of correlation between two sets of data (see Table \ref{tab:Pear_Spear}). In this case, they describe how likely it is that the observed slopes are actually caused by a statistical relationship between $\mathcal{S}$ and $\mathrm{\textit{N}_H}$. The Pearson coefficient is a measure of linear correlation and the Spearman coefficient is a measure of monotonic correlation. For nearly all of our log($\mathcal{S}$) vs log($\mathrm{\textit{N}_H}$) plots, both the Pearson and Spearman coefficient values had magnitudes that are close to zero ($|\rho_P|, |\rho_S| < 0.1$). This suggests that even though the data are consistent with a positive correlation, this correlation is very weak.

\begin{figure}
\centering
\includegraphics[width=0.85\columnwidth]{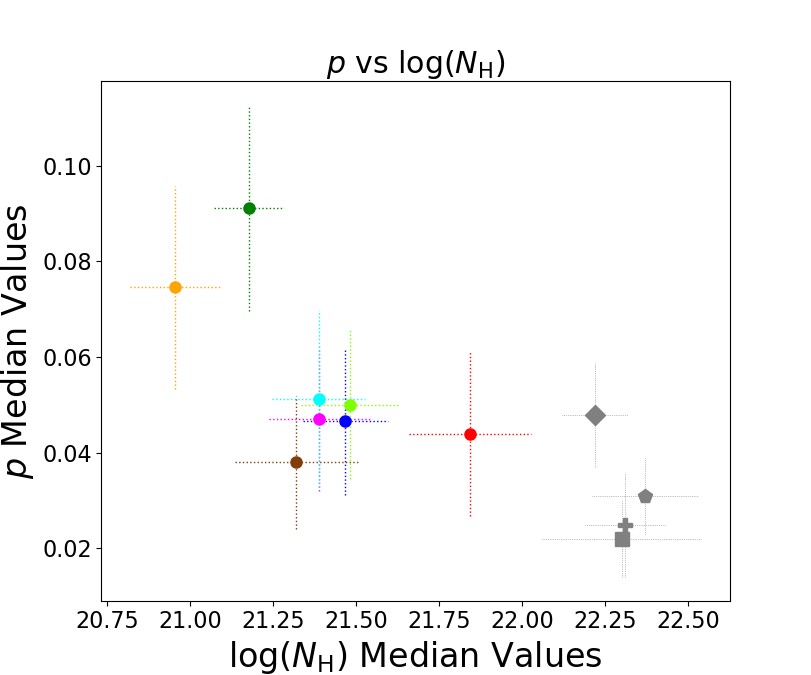}
\includegraphics[width=0.85\columnwidth]{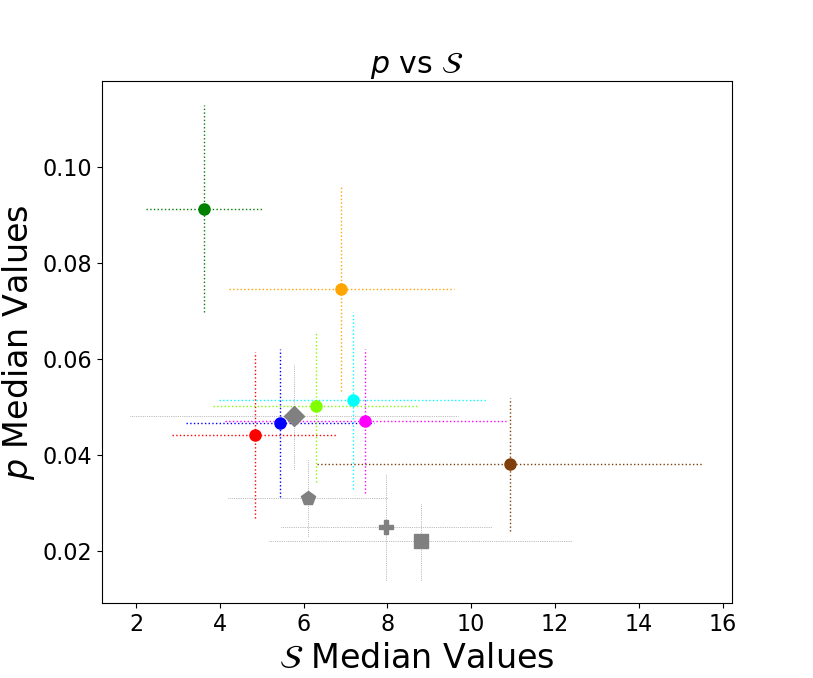}
\includegraphics[width=0.85\columnwidth]{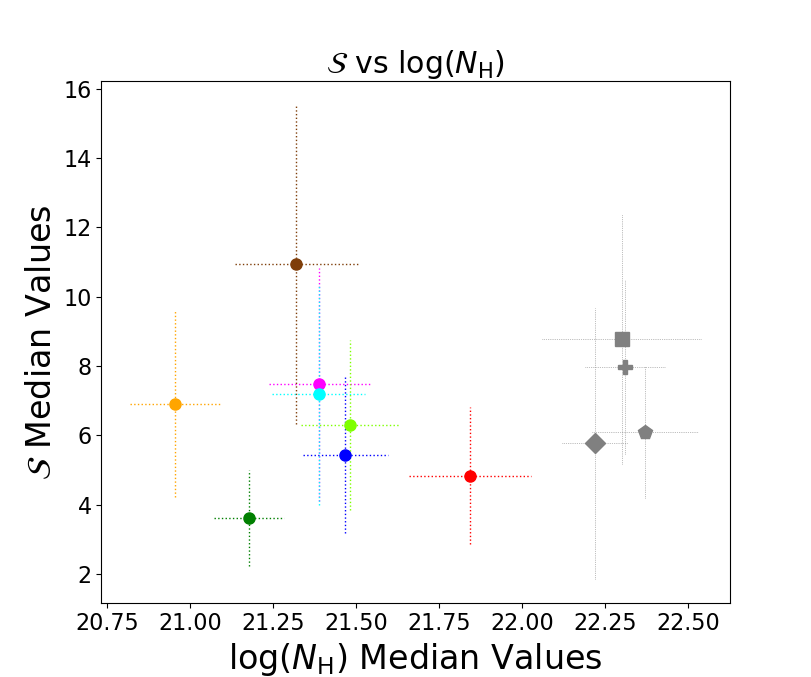}
\includegraphics[width=0.75\columnwidth]{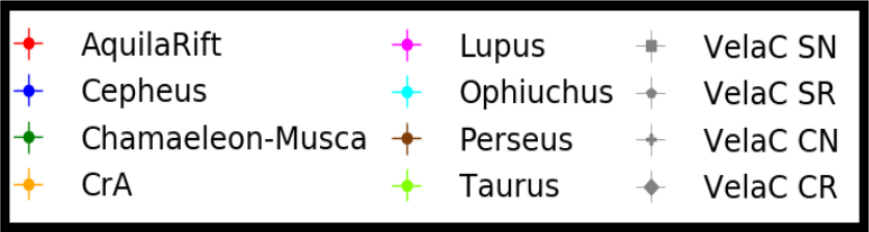} 
\caption{A set of plots comparing the median values of the three main polarization properties of our target clouds: polarization fraction (\textit{p}), dispersion in polarization angles ($\mathcal{S}$), and the log() of hydrogen column density (log$\mathrm{\textit{N}_H}$). The horizontal and vertical dotted lines represent the Median Absolute Deviation (MAD), which roughly indicates the distributions of each quantity. {\rev Dotted error bars are used} to represent MADs all throughout this paper. Here Vela\,C is broken into four sub-regions: South-Nest (SN), South-Ridge (SR), Centre-Nest (CN), and Centre-Ridge (CR).}
\label{fig:scatter}
\end{figure}
 
Overall, we were unable to establish a statistically robust relationship between $\mathcal{S}$ and $N_{\rm H}$. This lack of a measured correlation is in contrast to the results of \citet{PlanckXII}, which describe an observed trend of increasing $\mathcal{S}$ with increasing $N_{\rm H}$. The reason our data do not show this same correlation is likely a result of differences in the data samples. \citet{PlanckXII} combined all of their \textit{Planck} data and observed the average increase in $\mathcal{S}$ over a column density range of $1-20 \times 10^{21}~{\rm cm}^{-2}$, whereas all of our $\mathcal{S}$ vs $N_{\rm H}$ analysis was done on a per-cloud basis. Among our sampled clouds, the average range of column density values was only within  $0.98-11.0\times 10^{21}~{\rm cm}^{-2}$. In addition to our analyses covering a smaller range of column densities, the $\mathcal{S}$ vs $N_{\rm H}$ graph upon which \citet{PlanckXII}'s trend is based shows a region of little proportionality around the range of values in which our data generally falls.
Therefore, this discrepancy in ${\cal S}-N_{\rm H}$ correlation may be due to the limited coverage in column density range in our analyses.

\begin{figure*}
\begin{tabular}{c c}
\includegraphics[scale=0.408]{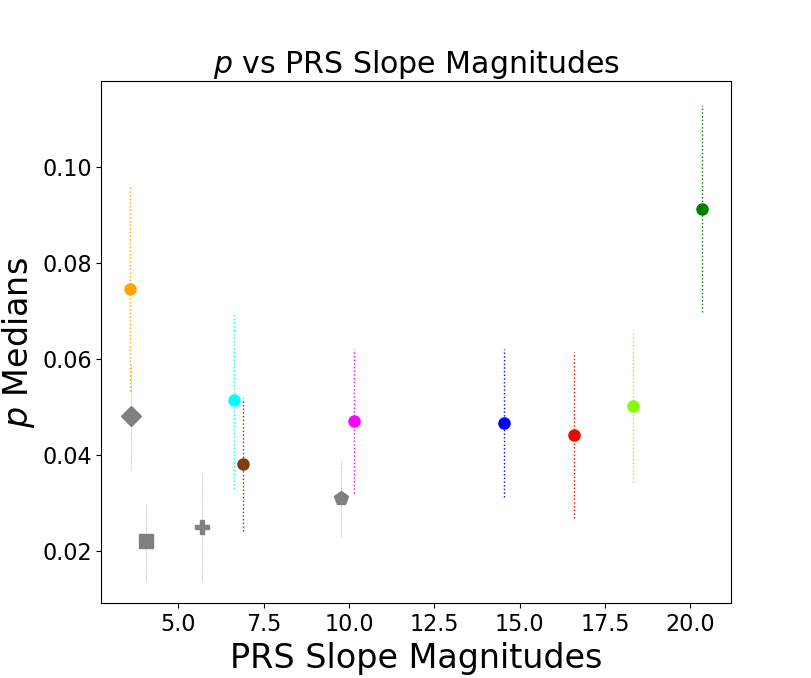} &
\includegraphics[scale=0.425]{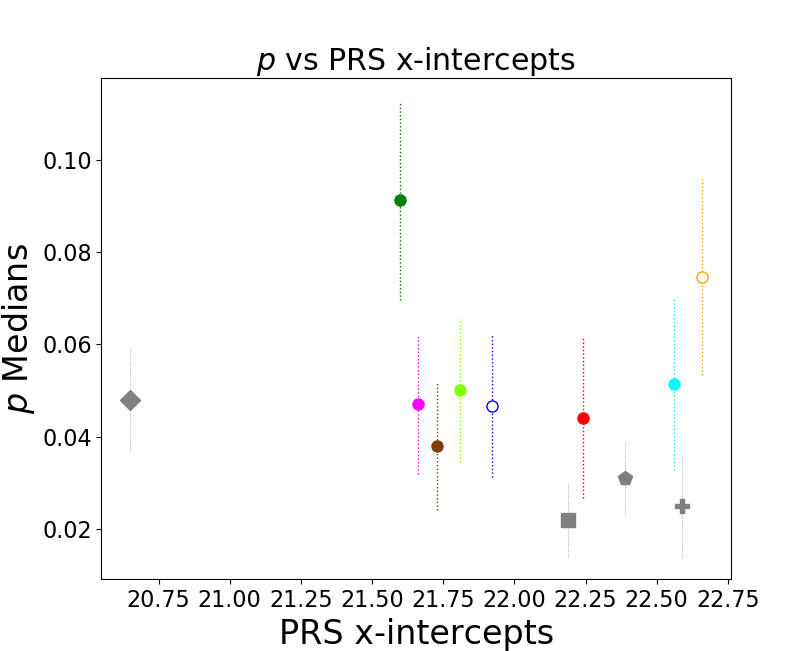} \\
\includegraphics[scale=0.427]{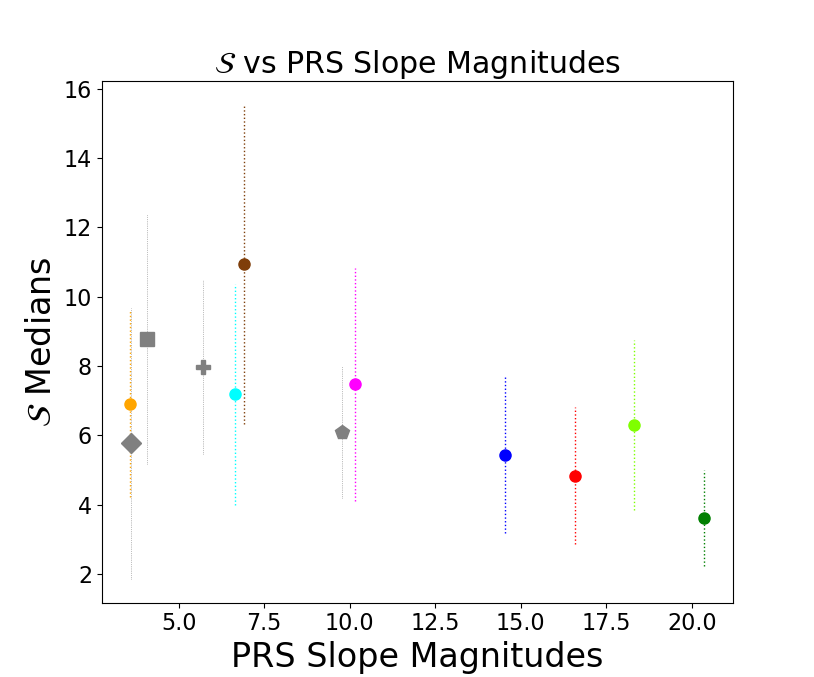} &
\includegraphics[scale=0.425]{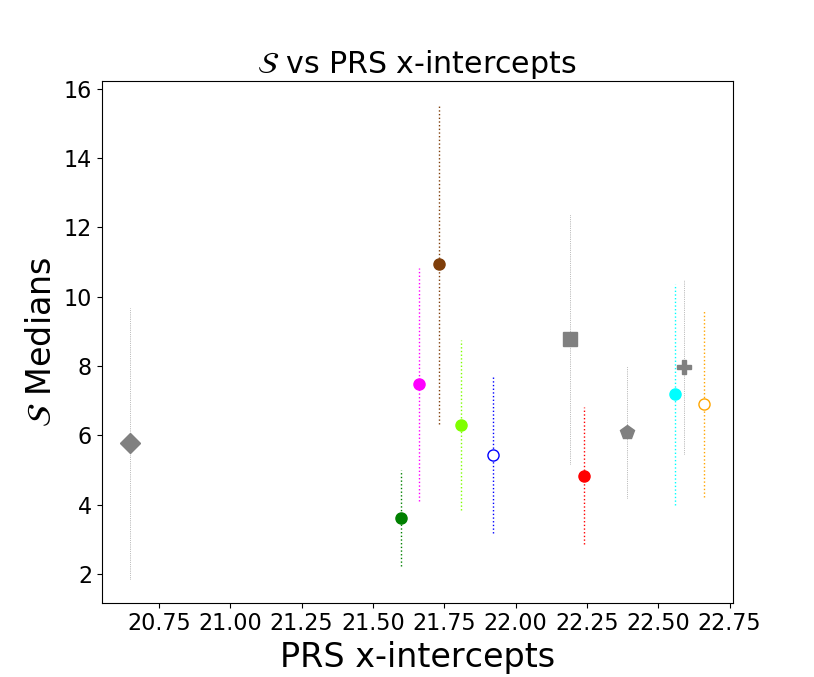}
\end{tabular}
\begin{tabular}{c}
\includegraphics[scale=0.4]{Plots/scatter_legend.png}
\end{tabular}

\caption{A series of scatter plots comparing the median \textit{p} and $\mathcal{S}$ values from our study to the slope and x-intercept values of the PRS plots, as determined by the study \citet{PRS}. Vertical bars are  again used to present each quantity's median absolute deviation (MAD).  We were unable to create similar error bars for the PRS slope or x-intercept values, as we could not obtain  any associated errors from \citet{PRS}. PRS values for Vela\,C as a whole were not obtainable as analysis of the total cloud was not conducted in the study. We therefore present one data point for each sub-region of the GMC:  the South-Ridge (SR), South-Nest (SN), Centre-Ridge (CR), and Centre-Nest (CN). Open circles indicate that the transition from positive to negative PRS was not observed by \citet{PRS}, so the x-intercept was extrapolated from the linear fit.}
\label{fig:scatter_PRS}
\end{figure*}

\subsubsection{Joint Correlations from Population Statistic Median Value Comparisons}
\label{sec:scatter}

To cross-compare typical polarization values for all clouds, we plot the median values of $p$, $N_{\mathrm{H}}$, and $\mathcal{S}$ as functions of one another in Figure \ref{fig:scatter}. Horizontal and vertical bars show the median absolute deviation (MAD) for each cloud in order to indicate the typical spread of the variable. 

 Note that in these scatter plots, we have divided our Vela\,C sightlines into four different sub-regions as first defined in the {\em Herschel} imaging survey of OB young stellar objects' (HOBYS) study of Vela\,C \citep{Hill}. Two sub-regions, the Centre-Ridge (CR) and South-Ridge (SR), show high column density filaments, while the Centre-Nest (CN) and South-Nest (SN) show extended lower column density filamentary structures at a variety of orientation angles. Of these sub-regions, the Centre-Ridge is the most active star forming region and includes a compact HII region. This HII region is powered by a cluster that contains an O9 star \citep{Ellerbroek2013, Hill}. It also has the highest polarization levels of all the sub-regions, and a high density structure that is oriented strongly perpendicular to the magnetic field as well \citep{Andersson2015, Soler2017,PRS}.

We see that the decrease of $p$~with increasing $N_{\mathrm{H}}$~and $\mathcal{S}$~observed within individual clouds is also seen in the comparison between median values for our sample of molecular clouds.  The presence of these trends on an inter-cloud scale, as opposed to a solely intra-cloud scale, suggests that these correlations between polarization quantities may be intrinsic, and not caused by cloud-specific properties or occurrences. Further investigation into this conclusion is left to future papers.

\subsection{Relative Orientation Analysis}
\label{sec:PRS_eval}

Another useful diagnostic of the magnetic field properties of a molecular cloud that is independent of our previous analysis is statistically comparing the orientation of the magnetic field to the orientation of cloud column density structures at every location on the map \citep{Soler2013}. In \citet{PlanckXXXV}, the authors showed that most of our sample clouds show a statistical change in alignment with respect to the magnetic field: lower column density sightlines have structures preferentially aligned with the magnetic field, while higher column density structures are more likely to have no preferred orientation or one that is perpendicular to the inferred magnetic field. When the relative orientation analysis was applied to {\tt RAMSES} simulations in \citet{Soler2013}, this change of relative orientation from parallel to perpendicular with increasing column density was only seen in high or intermediate magnetization simulations. \citet{PlanckXXXV} therefore argued that most of the ten nearby clouds in their study have the magnetic energy density equal to or larger than turbulence on cloud scales (see also \citealt{Chen2016}).

More recently, \cite{PRS} further quantified the relative orientation between magnetic field and gas structure using the projected Rayleigh statistics (PRS). In their analysis, \citet{PRS} calculated the PRS using the following equation: 
\begin{equation}
    Z_x = \frac{\sum_i^n \cos{\theta_i}}{\sqrt{n/2}},
    \label{eq:PRS}
\end{equation}
where $Z_x$ is the Projected Rayleigh statistic, and $\theta_i = 2\phi_i$ for which $\phi_i \in [-\frac{\pi}{2}, \frac{\pi}{2}]$, where $\phi$ is the relative angle between the  polarization orientation and $N_\textrm{H}$ gradient  orientation at each point. 
The sign of $Z_x$ in this application corresponds to different relative orientations of a cloud's magnetic field with respect to its column density gradients: $Z_x > 0$ indicates that the column density contours are preferentially oriented parallel to the magnetic field; $Z_x < 0$ instead indicates that the column density contours are preferentially oriented perpendicular to the magnetic field lines. $Z_x \approx 0$ represents a complete lack of measured preferential alignment between the two.
\cite{PRS} also fit a linear trend to $Z_x$~vs.~log($N_\mathrm{H}$) and reported, using the BLASTPol data of Vela\,C, that the PRS slope value and $N_\mathrm{H}$~intercept can be used to attempt to analyze the alignment between the magnetic field and column density variations.

Using the measured PRS slope magnitudes and $N_{\rm H}$ intercepts of the Planck clouds provided in \cite{PRS}, we plotted the correlation between the PRS measurements and the observed polarimetric properties in Figure~\ref{fig:scatter_PRS}.
Though a large slope often correlates with a more rapid shift in preferential alignment from parallel to perpendicular with increasing $N_\textrm{H}$ (thus a more ordered and stronger magnetic field), note however that the maximum amplitude of $Z_x$ is proportional to $n$ (particularly in the high-$n$ limit), where $n$~is the number of independent measurements of $\theta_i$. Therefore a cloud such as Corona-Australis, which has relatively few sightlines, will tend to show a lower range in $Z_x$~values, and thus a shallower slope compared to clouds with more sightlines. 

Nevertheless, we find a particularly noticeable trend in the $\mathcal{S}$ vs PRS slope magnitudes plot (Figure~\ref{fig:scatter_PRS}, bottom left). There seems to be a negative relationship between the dispersion in polarization angle $\mathcal{S}$ and the magnitude of slopes from linearly-fitting the PRS $Z_x$ vs log($\mathrm{\textit{N}_H}$) correlation. 
In short, large slope values tend to be observed in clouds with low average disorder in the projected magnetic field orientation, while small slope values are more often seen in clouds with more disordered magnetic fields \citep[see more discussions in][]{PRS}.

 This could be interpreted as both the ${\cal S}$ and PRS slope measurements being {\rev influenced by the} magnetic field strength and inclination angle of the clouds. A strong magnetic field will resist turbulent gas motions perpendicular to the magnetic field direction, and thus maintain a more ordered magnetic field orientation (lower $\mathcal{S}$), as opposed to a weak magnetic field in which turbulent gas motions will more easily be able to alter the magnetic field geometry (higher $\mathcal{S}$). Weak magnetic fields leading to a more disordered field geometry results in a lower degree of alignment with the cloud column density structure, and therefore lower magnitude PRS values.

Inclination angle of the magnetic field with respect to the plane of sky is another important factor in this relationship between $\mathcal{S}$ and PRS slopes, as it has a significant influence on each observable. As discussed in Section~\ref{sec:1Dplot}, a high inclination angle of the mean magnetic field will result in a high $\mathcal{S}$ value due to projection effects. This same exaggeration of the apparent magnetic disorder leads to a decrease in PRS slope magnitudes. 
In cases where the mean magnetic field is significantly inclined with respect to the plane of the sky, small variations in the magnetic field orientation caused by turbulence can result in large differences in projected magnetic field orientation. More disorder in the projected magnetic field orientation can result in less correlation with the orientation of cloud column density structure, thus creating a much weaker and more shallow $Z_x$~vs $N_{\mathrm{H}}$~trend.
\citet{Soler2013} also showed that in the rare case where the magnetic field is parallel to the line of sight, no preferential orientation with respect to column density can be seen.

To explore this trend further, we calculated the Pearson and Spearman coefficients of the median $\mathcal{S}$ vs PRS slope relationship. The result was a Pearson coefficient of $-0.771$ and a Spearman coefficient of $-0.741$, which suggests that these variables are correlated. Further analysis into the significance of this trend would be a valuable area of study.

As shown in \cite{PRS}, the values of $Z_x$ transition from positive to negative in the PRS plot for all but two clouds; this transition corresponds to a change in relative orientation from being preferentially parallel to preferentially perpendicular. The PRS x-intercept is a crude estimate of point at which this change occurs, and is based on the assumption that $Z_x$ and $N_{\rm H}$ share a linear relationship. Instead of x-intercepts naturally occurring within their PRS plots, Cepheus and Corona Australis had to have their the x-intercepts extrapolated from the linear fit \citep[see Fig.~A1 of][]{PRS}. These extrapolated values are indicated by hollow data points in Figure \ref{fig:scatter_PRS}. It has been suggested that this transition point signifies a shift in relative strength between gas kinetic energy density and magnetic energy density, which suggests a transition between being sub-Alfv\'enic and being super-Alfv\'enic \citep[also see e.g.,][]{Chen2016}. The PRS plots from \citet{PRS} seem to suggest that nearly all of the clouds in our sample thus fall into the categorization of trans-Alfv\'enic or sub-Alfv\'enic.

\section{Comparison with Synthetic Polarization Observations}\label{sec:Compare_king18}

\begin{figure}
    \centering
    \includegraphics[scale=0.32]{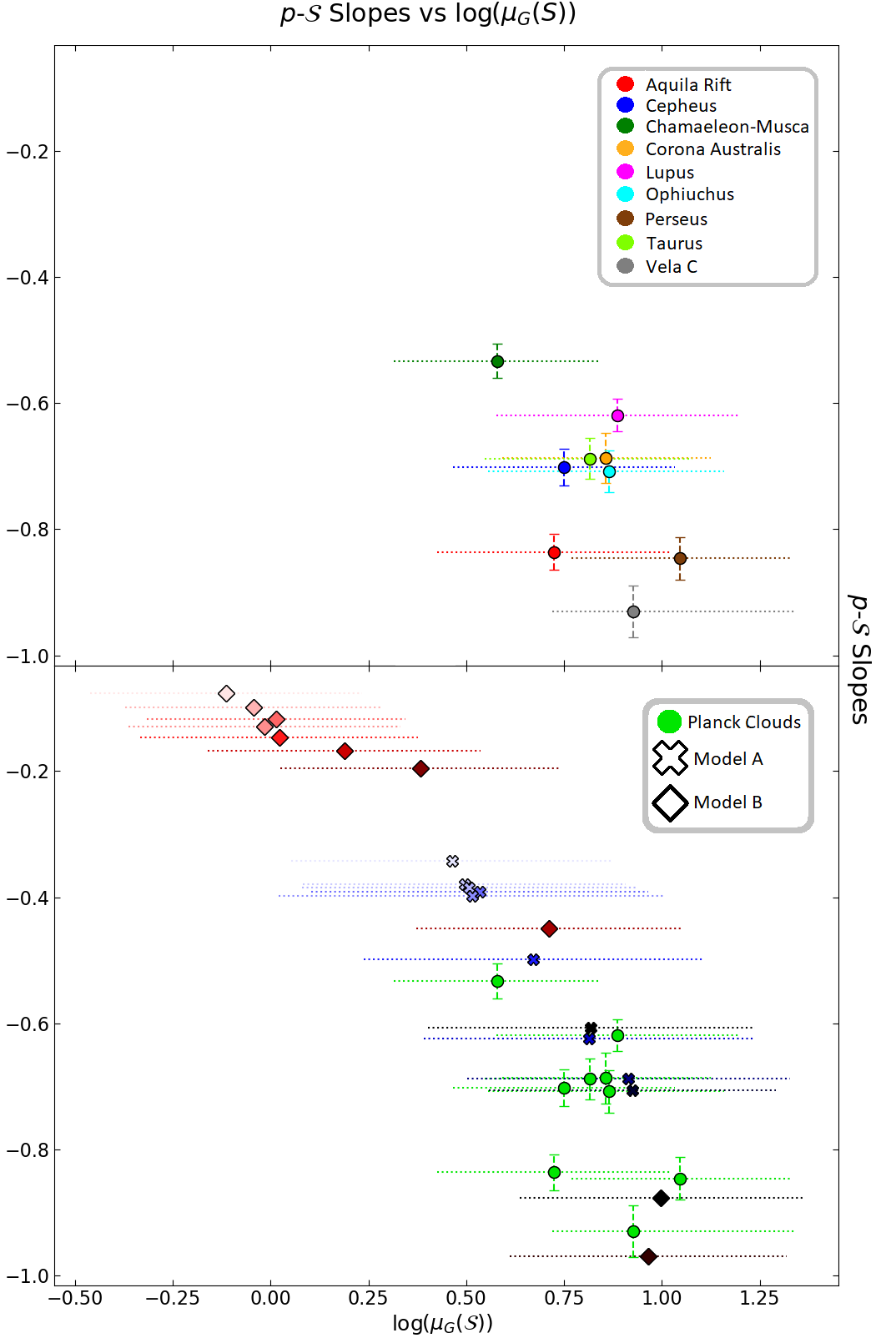}
    \includegraphics[scale=0.23]{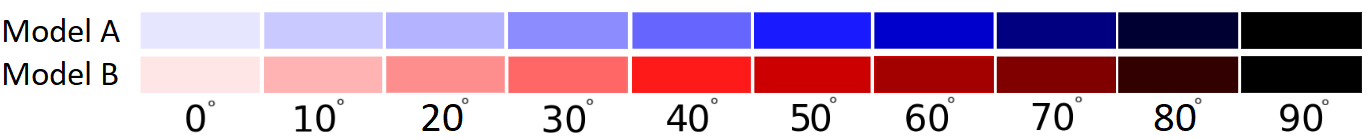}
    \caption{Correspondence between the slope of a cloud's $p$ vs $\mathcal{S}$ relationship and the geometric mean of ${\cal S}$ (in log scale), for both observed data ({\it top and bottom panels}) and those measured from synthetic observations (\citealt{king18}; {\it bottom panel}) of a more turbulent cloud (Model A; {\it thick plus}) and a more magnetically-dominated cloud (Model B; {\it diamonds}). The  average magnetic field inclination angles for the synthetic observations are represented by the color map depicted beneath the plot, with the inclination angle increasing from $0^\circ$ to $90^\circ$ in 10$^{\circ}$ intervals. The dotted horizontal lines on the x-axis show  the MAD of each cloud's $\mathcal{S}$ values and are thus a representation of the distribution of $\mathcal{S}$-values. The error bars on the y axis represent 3$\sigma$~confidence intervals of slope values obtained from bootstrap estimates.}
    \label{fig:pS_S}
\end{figure}

In Section \ref{sec:Continuous Plots} it was noted that Vela\,C had a strong negative correlation between its $p$ and $\mathcal{S}$ quantities. \citet{king18} attempted to explain this correlation in Vela\,C by comparing the observed polarization distributions to the polarization distributions of synthetic polarization maps made from 3D magnetohydrodynamic (MHD) simulations. In these simulations, molecular clouds are formed from the collision of two convergent gas flows, which creates a dense post-shock gas layer wherein filaments and cores form \citep{CO14, CO15}. Due to shock compression, the magnetic fields in these sheet-like clouds roughly align with the post-shock layer. Two simulations were examined: Model A and Model B. In Model A, the cloud formed in the post-shock region is relatively more turbulent and has higher turbulent-to-magnetic energy ratio, with an average Alfv\'{e}n Mach number $\mathcal{M}_{A,\mathrm{ps}}\approx 2.43$. In contrast, Model B was originally designed to simulate a local star-forming region within a magnetically-supported cloud (see \citealt{CO14, CO15}), and thus has well-ordered magnetic field structure with less-perturbed gas ($\mathcal{M}_{A,\mathrm{ps}} \approx 0.81$). These simulated clouds were ``observed'' at a variety of viewing angles with respect to the post-shock layer, and the only synthetic observations of these clouds that had similar $p$-$\mathcal{S}$ slopes to Vela\,C were those with large inclination angles of the magnetic fields with respect to the plane of sky ($\gamma \gtrsim 60^\circ$ for Model A and $\gamma \gtrsim 75^\circ$ for Model B). 

Here, we compare the polarization properties of nine clouds to the simulations presented in \citet{king18}. This allows us to better characterize the physical properties of the {\it Planck} clouds by directly comparing them with these synthetic observations and determining which synthetic observations (simulation models and viewing angles) can best reproduce the observed polarization properties within the clouds. This process also helps to validate the results of \citet{king18} by showing that the simulated observations share similarities with a larger variety of clouds than just Vela\,C. 

We note that $\mathcal{S}$~is a more useful quantity for comparisons than $p$, because \citet{king18} assumed uniform grain alignment and a prescribed coefficient of polarization fraction ($p_0$) of 0.15.  In fact, more recent work by \citet{King2019} showed that the $p$~vs.~$N_{\rm H}$~correlations can be strongly affected by non-homogeneous grain alignment efficiency in higher-column density regions; without a microphysically accurate grain alignment efficiency model, using this correlation to establish magnetic field properties of the target is a difficult proposition. The $p$~vs.~$\mathcal{S}$~correlations, however, show very little dependence on polarization efficiency assumptions \citep{King2019}. The mean $\mathcal{S}$ values and the $p$-$\mathcal{S}$ correlation power law index can therefore be considered to better reflect the magnetic properties of the clouds.

Figure \ref{fig:pS_S} shows a scatter plot of the $p$-$\mathcal{S}$ logarithmic slopes (power law indices; see Table~\ref{tab:Pear_Spear}) and the mean $\mathcal{S}$ values ($\mu_G({\cal S})$; see Table~\ref{tab:p_S}) for all of the {\it Planck} clouds considered in this study, Vela C, and  synthetic observations  of the two models discussed in \citet{king18}  viewed from different inclination angles with respect to the mean magnetic field direction in the post-shock layer.  We note that the synthetic polarization observations adopted here were analyzed at the full pixel-scale resolution of the simulations reported in \cite{king18}, which is significantly higher than the 0.6\,pc to 2.1\,pc of the {\em Planck} and BLASTPol maps of this study.  However, \citet{king18} found that the polarization distributions were not significantly affected by resolution, and we have verified with the 0.5\,pc resolution synthetic observations available from their study that the locations of the $p$-$\mathcal{S}$~vs mean $\mathcal{S}$~model points are not significantly affected by resolution. 
Also note that, \cite{king18} considered the {\it rotation angle} of the simulation box as the analysis parameter, however there are intrinsic angles of the average magnetic field relative to the shocked layer and thus the rotation axis \citep[see e.g.][]{Chen}. Here, we incorporated the intrinsic inclination angles of the cloud-scale magnetic field as measured in \cite{Chen} when varying the viewing angles of the synthetic observations, and thus the angles shown in Figure~\ref{fig:pS_S} represent the actual angle between the average magnetic field and the plane of sky.
For the {\em Planck}/BLASTPol data we have also attempted to estimate the uncertainty in our measurement of the $p$-$\mathcal{S}$~slope, through bootstrap errors. These bootstrap estimates were obtained by calculating the slope of each cloud 1000 times, with each slope estimated from a random sample of 25,000 sightlines.

We first note that our sample of clouds are not well matched by the synthetic polarization observations of the strong-field simulation, Model B. In order to match the mean $\mathcal{S}$~values of the observations with Model B, we would infer that the mean magnetic field direction is nearly aligned with the line of sight for 8 out of 9 clouds, which is not very likely. In addition, Model B predicts shallow slopes (less negative power law indices) in the $\log (p)$~vs $\log (S)$~plots when viewed from low-inclination lines of sight, which are not consistent with the slopes measured in our clouds. We therefore conclude that, in general, these nearby molecular clouds are not consistent with simulated clouds that have a highly-ordered magnetic field.

In contrast, Model A, the more turbulent cloud, can generally reproduce the mean $\mathcal{S}$ values, and  can better reproduce the $p$-$\mathcal{S}$ slopes for all clouds without requiring a single, shared magnetic field orientation for all clouds. Most clouds are consistent with synthetic observations with reasonable inclination angles ($\gamma \gtrsim 40^\circ$) of the magnetic field.

The exceptions to this are Perseus, Vela\,C, and the Aquila Rift, which appear to have a steeper $p$-$\mathcal{S}$ slope correlation than predicted from Model A.

We note that, as demonstrated in \citet{king18} and discussed in Section~\ref{sect:angles}, the $p$-$\mathcal{S}$ slope itself could be correlated with the cloud-scale magnetic field orientations with respect to the line of sight, which has a strong impact on the level of polarization fraction \citep[see e.g.,][]{Chen}. This indicates that the results shown in Figure~\ref{fig:pS_S} could be dependent on different methods for cloud sightline selection, as discussed in Appendix~\ref{appendix:A}. Nevertheless, all the different sightline masking methods we tested (see Appendix~\ref{appendix:A}) show better agreement with Model A than Model B for the observations. Indeed, a more detailed analysis with synthetic observations that mimic the effect of sightline masking is needed to use the $p$-$\mathcal{S}$~slope to infer the mean magnetic field inclination angle. This is beyond the scope of this paper.

 Among the observed clouds, Chamaeleon-Musca is the only one that could potentially be consistent with the magnetically-dominated Model B without requiring an unreasonably high inclination angle ($\gamma \gtrsim 75^\circ$) of the magnetic field. This is in fact the cloud with the highest polarization fraction in our analysis (see Table~\ref{tab:p_S} and Figures~\ref{fig:1D} and \ref{fig:scatter}). This lends credit to the idea that Chamaeleon-Musca may have a relatively strong cloud-scale magnetic field compared to the other clouds in this study.

Overall, our cloud polarization observation data in Figure \ref{fig:pS_S} better match the synthetic observations of a turbulent cloud that is not dynamically dominated by magnetic field (Model A in \citealt{king18}).
The better agreement with the more turbulent,  super-Alfv{\'e}nic Model A, would seem to conflict with the conclusions of the PRS analysis \citep{PRS} discussed in Section~\ref{sec:PRS_eval}, where the measured transition in relative gas-field orientation was taken to indicate that the gas must be trans- or sub-Alfv{\'e}nic at cloud scale.

 However, we note that the Alfv{\'e}n Mach~number of Model A cited from \citet{king18}, ${\cal M}_{\rm A} \approx 2.5$, is an average value among the entire cloud, and thus should not be taken as a diagnostic of all local gas conditions. \citet{king18} also did not consider the gas flow direction with respect to the magnetic field when calculating $\mathcal{M}_\mathrm{A}$. In fact, the value of $\mathcal{M}_\mathrm{A}$ would be reduced if only the velocity component perpendicular to local magnetic field is considered. Under this definition, the simulated cloud in Model A of \citet{king18} is indeed trans-Alfv{\'e}nic (see also \citealt{Chen} for more discussions). We would also like to point out that the derived Alfv{\'e}n Mach~numbers within individual clouds should only be considered as references, not definitive properties of the entire clouds. In fact, since molecular clouds are spatially large and likely cover a wide range of physical environments, it is inappropriate to use a single value to represent the properties of the entire cloud.

\section{Conclusions}\label{sec:Conclusions}

The goal of this study was to characterize the magnetic field properties of nine nearby molecular clouds. This characterization was done by comparing 353 GHz polarization data of eight clouds from the {\em Planck} survey and 500~$\mu$m polarization data on Vela\,C from BLASTPol, and investigating polarization observables such as polarization fraction ($p$) and the local dispersion in polarization angles ($\mathcal{S}$). We also examine the correlation of these polarization properties with \textit{Planck} hydrogen column density maps of our target regions. Comparisons were drawn between our observations and those of \citet{PlanckXX, PlanckXII, PlanckXXXV} in particular. We also compare our observations to synthetic polarization observations of two simulations from \citet{king18}: Model A, a more turbulent simulation where the energy density in the magnetic field is comparable to the energy density of turbulent gas motions, and Model B, a simulation where the magnetic energy density dominates turbulence with a very ordered magnetic field. The main conclusions of our paper are as follows:

\begin{enumerate}[label=\arabic*)]

\item Using the methods described in \citet{Chen}, we estimate the average inclination angle of each cloud's magnetic field. This process is based on each cloud's maximum polarization fraction value ($p_{\rm max}$) and the 1D probability distribution function of $p$~that are shown in Figure \ref{fig:1D}. The estimated inclination angles that we obtained are presented in Table \ref{tab:inc}, and range from $17^\circ$ to $68^\circ$.

\item In \citet{king18}, it is suggested that the slope of a cloud's $p$ vs $\mathcal{S}$ relationship is strongly affected by the inclination angle of the cloud's magnetic field with respect to the plane of sky. We were able to provide support for this assertion in Figure \ref{fig:pS_S} by plotting each cloud's $p$ vs $\mathcal{S}$ slope, including the values for \citet{king18}'s Model A and Model B at  $10^\circ$ increments between $0^\circ {\rm and }~90^\circ$, against its geometric mean $\mathcal{S}$ value ($\mu_G({\cal S})$). We find an increase in $p$ vs $\mathcal{S}$ slope magnitude with increases in $\mathcal{S}$, which is in turn correlated with increases in inclination angle (see Section~\ref{sect:angles}). This suggests that as magnetic fields become more apparently disordered, by either a decrease in field strength or an increase in inclination angle, the rate at which $p$ will vary with respect to $\mathcal{S}$ increases.

\item In most of our cloud sample, we observe a systematic trend of decreases in $p$ with increases in $N_{\rm H}$. This trend is present both within and between clouds, with the exception: Chamaeleon-Musca, Lupus, and Taurus. These clouds do not show strong correlations between $p$ and $N_{\rm H}$, as evidenced by their very low Pearson and Spearman coefficients as well as Chamaeleon-Musca's low slope value and extremely low Pearson and Spearman coefficients. It is possible that the difference lies in their narrow $N_{\rm H}$ distributions. We also note that these clouds have fairly low median $N_{\rm H}$ values. The small range of $N_{\rm H}$ values in these clouds may not be providing a wide enough range of values for a significant trend to be observed. In their analysis, \citet{king18} compared the $p$ vs $N_{\rm H}$ slopes from their synthetic polarization observations of two different colliding flow simulations with that of their Vela\,C BLASTPol observation, and they found that their simulations were unable to accurately recreate the observed drop in $p$ vs $N_{\rm H}$. We compare our observed slope values and find the same result: neither Model A nor Model B has a $p$ vs $N_{\rm H}$ relationship that resembles those derived from the {\em Planck} or BLASTPol data.

\item We were unable to establish a statistically significant relationship between hydrogen column density ($N_{\rm H}$) and dispersion in polarization angles ($\mathcal{S}$). Although nearly all eigenvalue slopes produced by our covariance matrices showed the predicted positive trend, their associated Pearson and Spearman coefficients were incredibly low (Table~\ref{tab:Pear_Spear}), and thus the trend is not statistically significant. \citet{PlanckXII} reported a trend of increasing $\mathcal{S}$ with $N_{\rm H}$, but we were not able to confirm its existence with the analytical methods used in this work.

\item A negative correlation between dispersion in polarization angles ($\mathcal{S}$) and the magnitude of projected Rayleigh statistic (PRS) slope magnitudes has been observed for all clouds. The statistical relevance of this correlation has been evaluated through the use of Pearson and Spearman coefficients, and it is found to be significant. We believe that this relationship is caused by the two variables' individual dependencies on both magnetic field strength and viewing geometry, and may be useful in future attempts to determine the inclination angle and relative energetic importance of the magnetic field vs turbulent gas motions within molecular clouds.

\item Among the clouds observed,  Chamaeleon-Musca shows the highest $p$~values, the lowest $\mathcal{S}$~values, and the shallowest $p$ vs $\mathcal{S}$ slope magnitude.  Chamaeleon-Musca is also the only cloud with $p$-$\mathcal{S}$ slope and mean ($\mathcal{S}$) values that are consistent with those of \citet{king18}'s strongly magnetized molecular cloud simulation, Model B. This could imply that the magnetic field in Chamaeleon-Musca is more ordered, and possibly more dynamically significant, than the magnetic fields in other molecular clouds considered in this study.
However, the estimated inclination angle with respect to the plane of the sky of the cloud-scale magnetic field in Chamaeleon-Musca is  $\approx 17^\circ$, which is the lowest of all clouds in our sample. A low inclination angle of the mean magnetic field would also result in low $\mathcal{S}$ values and a shallow $p$-$\mathcal{S}$ slope, and so with these observations alone we can not conclusively state that Chamaeleon-Musca has a stronger magnetic field than the other clouds in our sample.
\end{enumerate}

It is our intent that these data be used for future research on the effects that magnetic fields have on star formation. For future analysis of these clouds, their polarization parameters, and the relation of these results to rates of star formation, we suggest that clouds such as Perseus be broken down into their sub-regions, as we have done for Vela\,C in parts of this paper. Our whole-cloud analysis over-simplifies the complicated magnetic field morphology of these molecular clouds and assumes that the inclination angle of the magnetic field is more or less constant across each map.  In addition, future synthetic observations should attempt to replicate the biases associated with sightline selections, and polarized background and foreground emission. These selections affect the precise values of the average properties and correlations of polarization measurements. Future work would also benefit from analyses using more tracers of magnetic field properties, including observations of background stars in optical or near-infrared bands, Faraday rotation, and Zeeman splitting. To further understand the observational biases towards highly polarized regions that is caused by polarization intensity-based sightline selection criteria, it would be helpful to apply a similar selection criteria to synthetic polarization maps by masking regions of low polarization intensity. This simulation masking may help to determine how these sightline selection criteria affect their resulting polarization distributions.

\section*{Acknowledgements}
This work was supported by the University of Virginia and the National Radio Astronomy Observatory, including its funders (AUI and the NSF). This research made use of Astropy,\footnote{http://www.astropy.org} a community-developed core Python package for Astronomy \citep{astropy:2013, astropy:2018}. CHS acknowledges partial summer support from the University of Virginia through the Virginia Initiative on Cosmic Origins. LMF, CYC and ZYL are supported in part by NSF AST-1815784. LMF and ZYL acknowledge support from NASA 80NSSC18K0481. ZYL is supported in part by NASA 80NSSC18K1095 and NSF AST-1716259 and 1910106. This work has also been partially supported by NASA NNX13AE50G through the BLAST collaboration. PKK is supported by a Livermore Graduate Scholarship at Lawrence Livermore National Laboratory, and acknowledges the Jefferson Scholars Foundation for additional support through a graduate fellowship. Part of this work was performed under the auspices of the Department of Energy by Lawrence Livermore National Laboratory under Contract DE-AC52-07NA27344. LLNL-JRNL-785178. Colin Sullivan was a summer student at the National Radio Astronomy Observatory (NRAO). LMF was supported by a Jansky Fellowship from NRAO and also by funding from the Natural Sciences and Engineering Research Council of Canada (NSERC).

\section*{Data Availability}
{\rev No new data were created or analysed as part of this research. The simulation data underlying this article will be shared on reasonable request to P. K. King and/or C.-Y Chen.
}

\bibliographystyle{mnras}
\bibliography{references}

\appendix

\section{Discussion of {\em Planck} Sightline Selection Biases}\label{appendix:A}

In analyzing polarization data in this paper we have only included sightlines above a column density threshold characteristic of the diffuse ISM at the same Galactic latitude, and required that the polarized intensity $P$~be at least $3\times$~larger than the associated uncertainty ($P\,\geq\,3\sigma_{P}$) in the {\em Planck} maps (see Section \ref{sec:masking} for more details). The goal of the column density masking threshold is to only analyze sightlines that are above the typical diffuse ISM background column density, and therefore likely associated with the cloud. The goal of the polarization selection criteria is to only analyze polarization data that has a high degree of statistical significance, and therefore does not require complicated error debiasing analysis as discussed in \citet{PlanckXIX}. 
However, it is important to examine whether these selection criteria bias our fits of the relationships between polarization measureables by rejecting regions of the cloud where the polarization is weak.

\begin{figure*}
\includegraphics[scale=0.8]{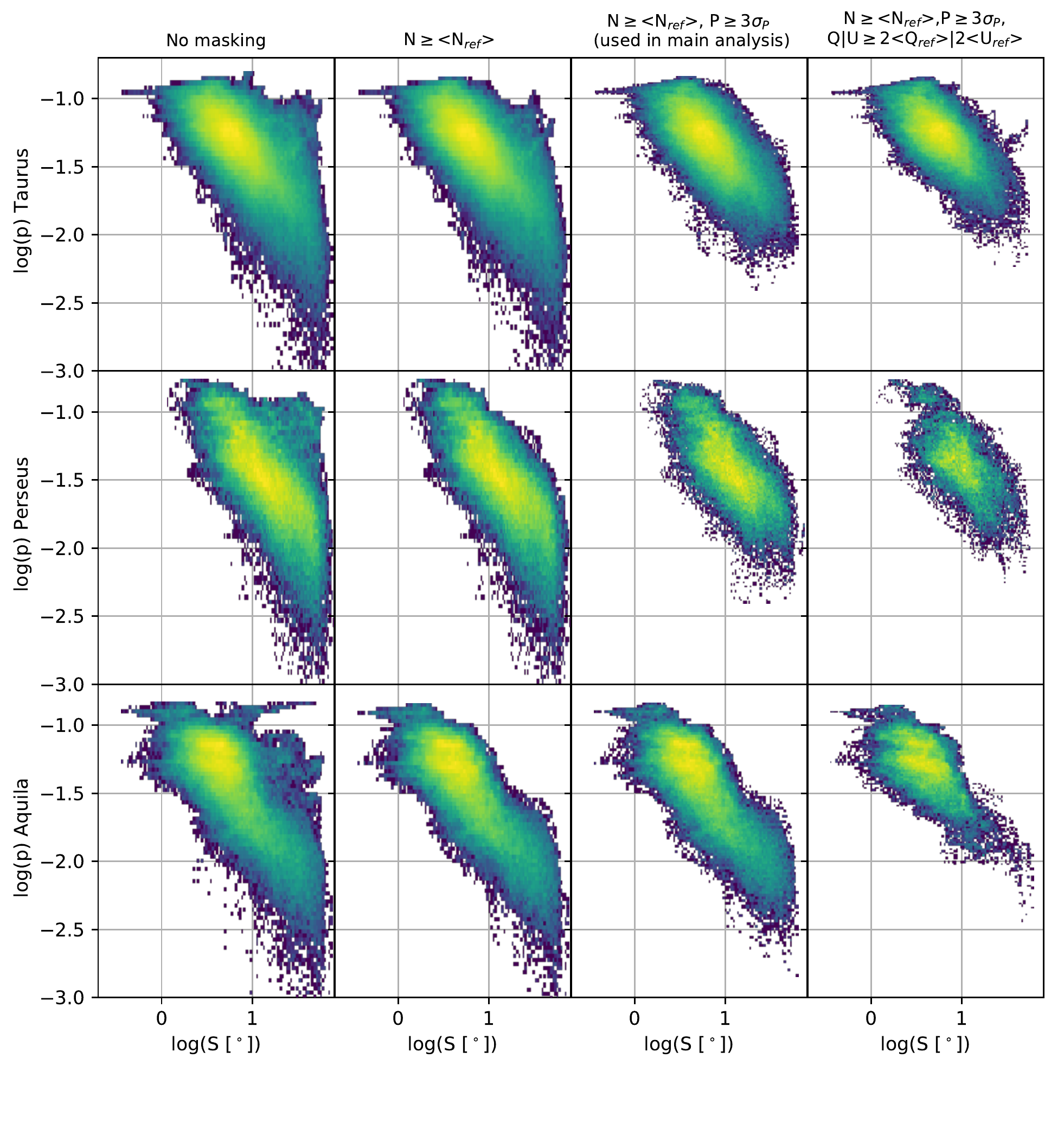}
\caption{ 2D histograms of the polarization fraction ${\rev log}(p)$~and distribution of polarization angles ${\rev \log}(S)$ for three of our {\em Planck} polarization maps: Taurus (top row), Perseus (middle row), and Aquila Rift (bottom row).  Each of the columns shows how the $p$~and $S$~distributions change for different choices of the masking criteria.  The left most column shows the distribution when no sightlines are masked, and the next column shows the distribution when the only masking criteria is that the sightlines used in the analysis must have a column density $N_{\mathrm H}$ that is greater than the average column density in a diffuse ISM field at the same Galactic latitude. The third column requires that the polarized intensity be at least a $3\sigma$\,detection in addition to the column density threshold (this is what is used for the analysis in the main text of the paper).  The rightmost column also requires that for each sightline used in the analysis, either $Q$~or~$U$~must be at least 2 times as large as the RMS value of Q and U from the reference diffuse ISM field at the same Galactic latitude, which is the selection criteria used in \citet{PlanckXXXV}.}
\label{fig:pS_masking}
\end{figure*}

\begin{figure*}
\includegraphics[scale=0.8]{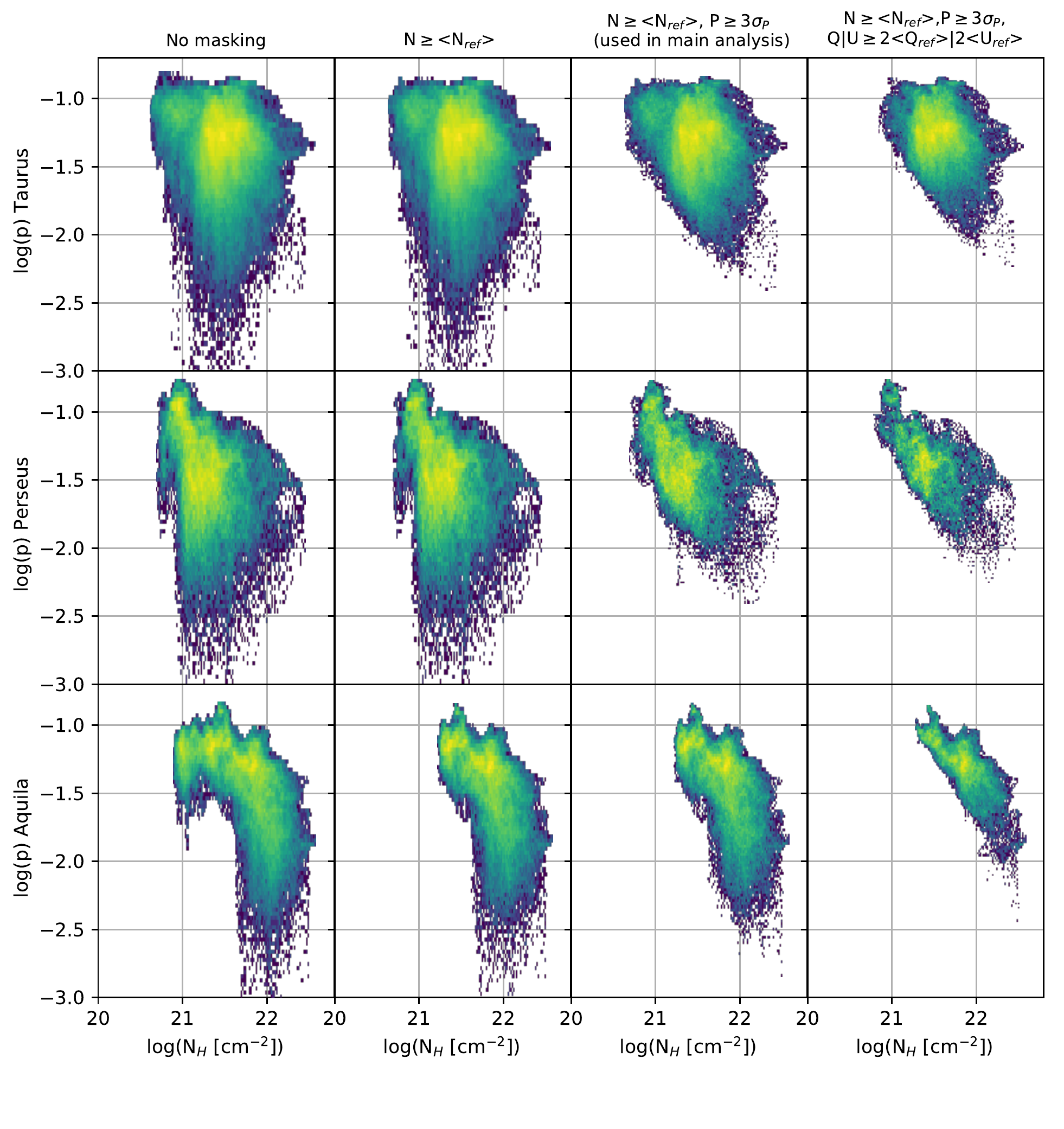}
\caption{ 2D histograms of the polarization fraction log($p$)~and log($N_{H}$) for three  of our  {\em Planck} polarization maps: Taurus (top row), Perseus (middle row), and Aquila (bottom row).  Each of the columns shows how the $p$~and $(N_{H})$~distributions change for different choices of the masking criteria, as described in the caption for Figure \ref{fig:pS_masking}.}
\label{fig:pN_masking}
\end{figure*}

 In Figures \ref{fig:pS_masking} and  \ref{fig:pN_masking} we examine the distribution of $p$ vs $\mathcal{S}$~points for the Taurus, Perseus, and Aquila maps using four different masking methods.  In the left-most panel, the distribution of all sightlines with no masking shows more points at low $p$ and high $\mathcal{S}$ as well as more points at high $p$ and high $\mathcal{S}$ compared to the third column, which shows the sightlines for the masking criteria used in the main paper text. The high $p$ and high $\mathcal{S}$  sightlines are absent in the second column, where we removed sightlines where the column density $N_{\mathrm H}$~ is less than the mean column density in a diffuse ISM region at the same Galactic latitude. This is not surprising, as low column density sightlines tend to have higher polarization fractions for the same values of $p$ and $\mathcal{S}$ \citep{Fissel}.  Masking these diffuse sightlines has the effect of making the correlation between log$(p)$~and $\mathcal{S}$~closer to a linear trend.

When we apply the second criteria for selecting the sightlines used in the main paper text, that the polarized intensity $P$~be at least 3$\times$~larger than the $\sigma_P$, we find that this removes the lowest-$p$ values in the plots, which tend to have high $\mathcal{S}$-values. Detections of polarization below this level are not statistically significant, but this does bias our data slightly towards lower average polarization angle dispersion $\mathcal{S}$, and higher average fractional polarization values $p$.

\begin{figure}
\includegraphics[scale=0.6]{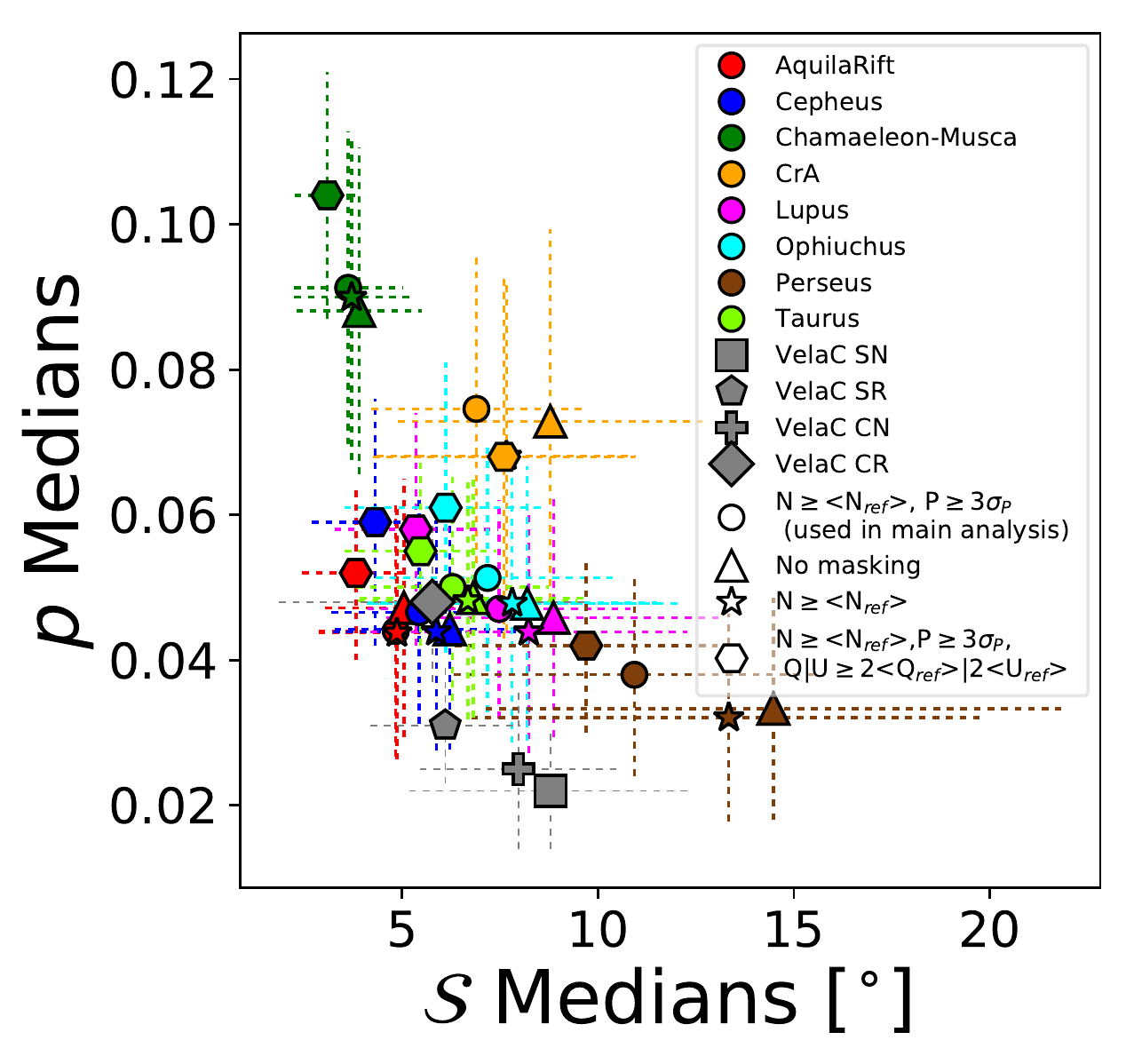}
\includegraphics[scale=0.6]{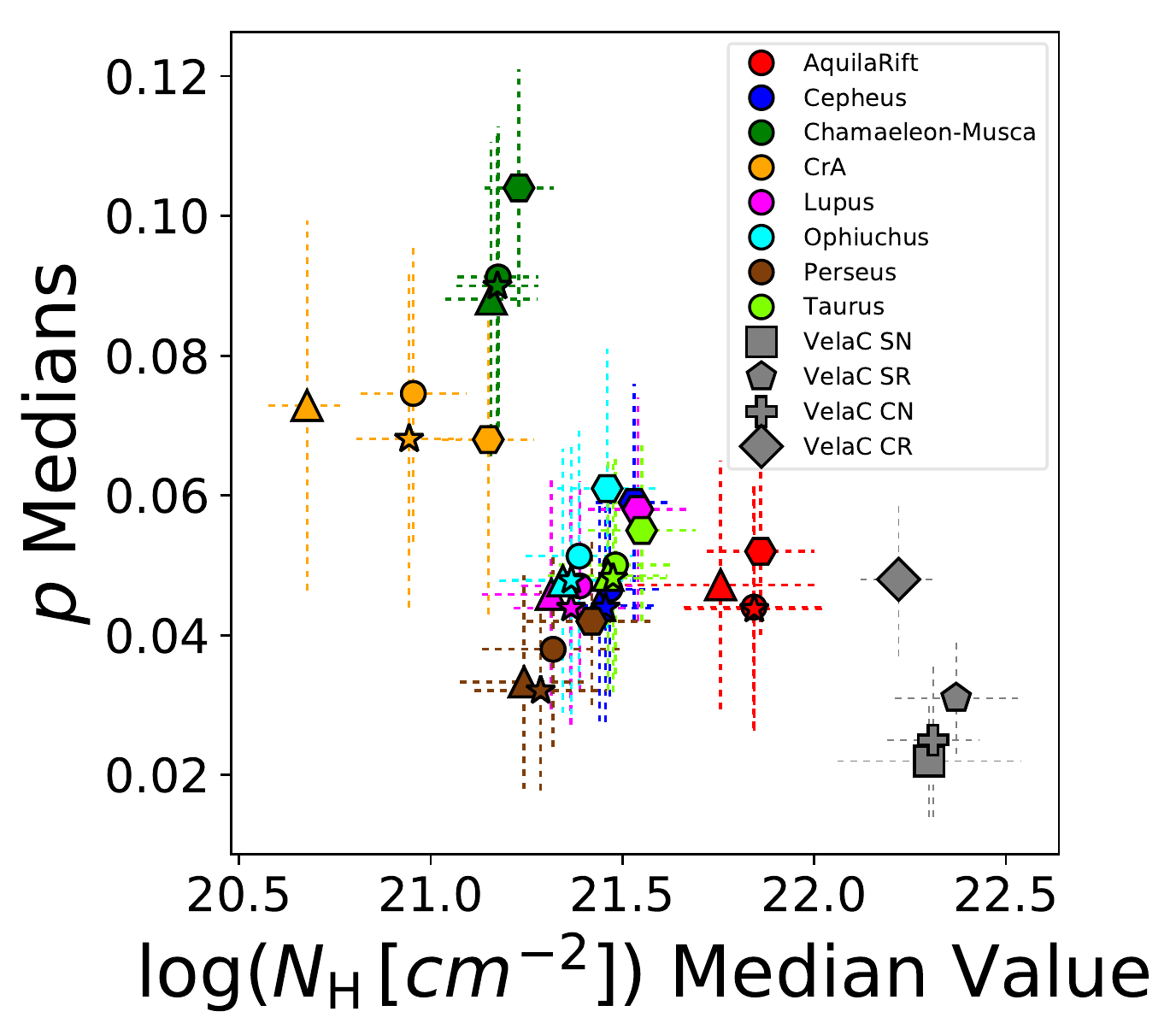}
\caption{ Comparison of the median polarization fraction $p$ with  polarization angle dispersion $\mathcal{S}$ (top panel) and the logarithm of hydrogen column density log$(N_{\mathrm H})$ (bottom panel). This figure is similar to Figure \ref{fig:scatter_PRS}, except that for the {\em Planck} observed clouds we show the median values using the four different masking criteria discussed in Appendix \ref{appendix:A}.  The dotted lines indicate the median absolute deviations (MAD) for each quantity, and are intended to indicate the range of polarization values for each cloud.}
\label{fig:pNSmeds}
\end{figure}

\begin{figure} 
\includegraphics[scale=0.6]{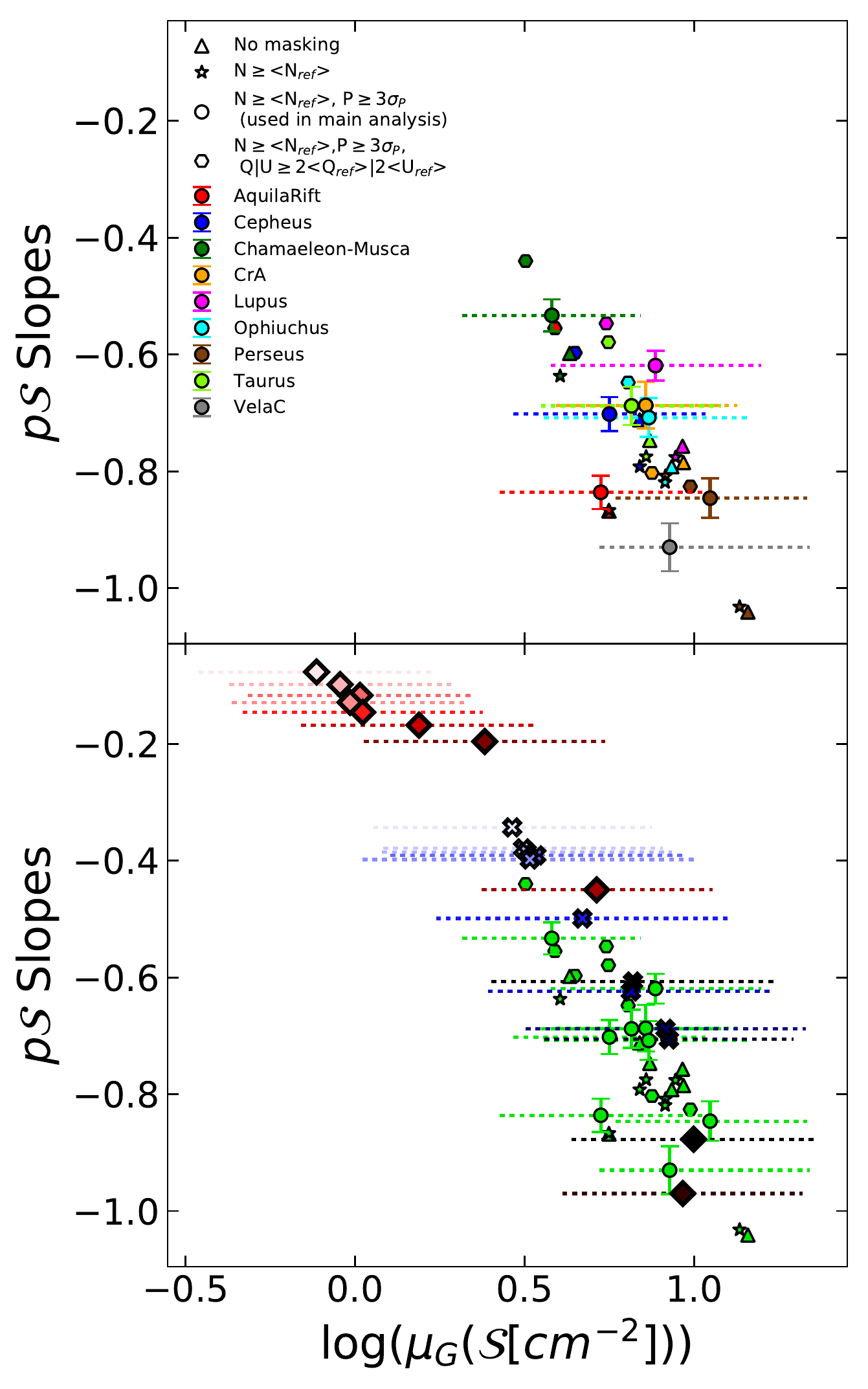}
\caption{ Plot of the fitted log($p$) vs log($\mathcal{S}$) slope index compared to the logarithm of geometric mean polarization distribution $\mathcal{S}$, similar to Figure \ref{fig:pS_S}, but comparing different masking strategies for the {\em Planck} data.  In addition to the masking strategy used in the main paper text as described in Section \ref{sec:masking} (circles, $N_{\mathrm H}$ larger than the RMS $N_{\mathrm H}$ of a reference diffuse ISM field at the same Galactic latitude, and $P\,\geq\,3\sigma_{P}$) we also show $p$-$\mathcal{S}$~slope vs $\mu_G(S)$~with no masking applied (triangles), and column density threshold masking only (stars). We also show $p$-$\mathcal{S}$~slope vs $\mu_G(\mathcal{S})$~ using the same masking as is used in the main paper, but with the additional requirement that either $Q$~or $U$~be at least twice as large as the RMS $Q$~or $U$~of the diffuse ISM reference field as required by \citet{PlanckXXXV} (hexagons). We show 16th and 84th percentiles of the $\mathcal{S}$~distribution (dotted horizontal lines) only for the models and standard masking used in the main paper. The vertical error bars are the 3$\sigma$~confidence intervals for the $p$-$\mathcal{S}$~slope fits derived from bootstrapping analysis described in Section \ref{sec:Compare_king18}.}

\label{fig:pS_vs_Smasking}
\end{figure}

To be fully consistent with \cite{PlanckXXXV}, which used the same Planck maps as this study but at 10$\arcmin$~FWHM resolution instead of 15$\arcmin$~FWHM resolution, we would have to apply one additional sightline selection criteria. This selection criteria would require that the polarized radiation be at least 2$\times$~as bright as the emission in the same reference diffuse ISM field used to set the column density threshold.  More specifically, the additional criteria for each sightline is that either
\begin{equation}\label{eqn:qref}
|Q|\,\geq 2 |Q_{ref}|,
\end{equation}
or 
\begin{equation}\label{eqn:uref}
|U|\,\geq 2U_{ref},
\end{equation}
where $Q_{ref}$~and $U_{ref}$~are the RMS $Q$~and $U$~values in the reference diffuse ISM field.  However, for many clouds this cut eliminates most of the sightlines (see for example Aquila in Figure \ref{fig:pS_masking}).  It also leads to the rejection of most low-$p$/high-$\mathcal{S}$~sightlines, which further biases the sample to lower $\mathcal{S}$~values and shallower log$(p)$~vs $log(\mathcal{S})$~slope indices.

Similarly, the same masking criteria also affect the relative distribution of polarization fraction $p$~and column density ($N_{\mathrm H}$), as shown in Figure \ref{fig:pN_masking}.  Applying the column density threshold criteria removes mostly high $p$~sightlines. Applying the requirement that $P\,\geq\,3\sigma_{P}$, removes mostly low $p$~sightlines, but tends to remove somewhat higher $p$-values at lower column densities (e.g. for Taurus and Perseus).  Applying the \cite{PlanckXXXV} requirements from Equations \ref{eqn:qref} and \ref{eqn:uref} removes even more low-$p$~sightlines and tends to make the log$(p)$~vs~log$(S)$~slope significantly more shallow.

We can also check whether the sightline masking criteria affect the average polarization properties of the clouds.  In Figure \ref{fig:pNSmeds}, we show that the choice of sightline selection method certainly affects the median values of $p$, $\mathcal{S}$, and ${\rev \log}(N_{\mathrm H})$, but it doesn't typically change the relative trends between the clouds.  The clouds that have the highest average polarization fraction with the sightline selection method used in the main paper text (Corona Australis and Chamaeleon-Musca), also have the highest polarization fraction if no sightline masking is applied, or if more aggressive polarization intensity masking is applied. Similarly, clouds that have high polarization angle dispersions, like Perseus, have high dispersion values for all of our tested sightline selection methods.

Finally, we discuss the effect of sightline selection on Figure \ref{fig:pS_S}, where we compared the $p$~vs.~$\mathcal{S}$~slope as a function of the geometric mean of $\mathcal{S}$, for the {\em Planck}-observed polarization maps, the BLASTPol polarization map of Vela\,C, and the synthetic models of the {\tt Athena} MHD models presented in \cite{king18}.  
We argue that the mapping method used in the main text should give the best comparison dataset for the simulations from \cite{king18}, as the synthetic observations only integrate over voxels from the post-shock region, which would be equivalent to the dense molecular cloud regions, and which has extremely high signal to noise polarization data.

Figure \ref{fig:pS_vs_Smasking} shows how the distribution of $p$-$\mathcal{S}$ slope vs mean $\mathcal{S}$ changes with sightline selection criteria. While the absolute values of the $p$-$\mathcal{S}$ slope do indeed change as more masking criteria are applied, the relative trends between the clouds are not significantly affected. Furthermore, all masking methods still show better agreement with the more turbulent Model A, rather than the highly sub-Alfv\'enic Model B. However, using the comparison shown in Figure \ref{fig:pS_S} and Figure \ref{fig:pS_vs_Smasking} to constrain the inclination angle of the magnetic field will require a more careful accounting for the sightline selection effects when comparing with synthetic observations, which is beyond the scope of this work.

\bsp
\label{lastpage}
\end{document}